\documentstyle[stwol,epsfig]{article}

\input{psfig}

\def\be{\begin{equation}}
\def\ee{\end{equation}}
\def\bea{\begin{eqnarray}}
\def\eea{\end{eqnarray}}
\def\as{\alpha_s}
\def\pom{{I\!\!P}}                % gives pomeron symbol

\def\prop{\sim}
\def\gev{\ {\rm GeV}}
\def\gev2{\ {\rm GeV}^2}
\begin{document}

\title{TESTS OF QCD AT LOW $x$}

\author{ H. ABRAMOWICZ }

\address{School of Physics and Astronomy\\ 
Raymond and Beverly Sackler Faculty of Exact Sciences\\ 
Tel Aviv University, Israel}

\twocolumn[\maketitle
\vspace{-5cm}
\begin{flushright}
TAUP 2396-96 \\ 
November 1996 \\
\end{flushright}
\vspace{4cm}
\abstracts{This talk reviews the latest measurements 
of the proton, Pomeron and photon structure functions. These
measurements, especially at low $x$ and/or low $Q^2$ lead to new
insight into the picture of hadronic interactions.  }
]

\section{Introduction}

One of the big challenges within Quantum Chromodynamics (QCD) as the
theory of strong interactions is the understanding of its role in
describing the approach to the non-perturbative, soft interactions, at
least at a qualitative level if not quantitative.

The soft hadron-hadron interactions are well described by the
Regge phenomenology in which the interaction is viewed as due to
exchanges of Regge poles. The Regge poles can be classified into
different families according to their quantum numbers. The Regge
poles with quantum numbers of mesons form linear trajectories in
the $m^2, l$ plane, where $m$ is the mass of the meson and $l$ its
spin. The continuation of a trajectory to negative values of $m^2$
leads to a parameterization in terms of $t$, the
square of the four momentum transfer, as follows:
\begin{equation}
\alpha(t) = \alpha_0 + \alpha' \cdot  t \, ,
\label{eq:alpha}
\end{equation}
where $\alpha_0$ is the intercept and $\alpha'$ is the slope of the
trajectory.  Among all possible families of Regge poles there is a
special one, with the quantum numbers of the vacuum, called the
Pomeron ($\pom$) trajectory. There are no known bound states lying on
this trajectory (glueballs would be expected to form this trajectory)
and its parameters have been determined
experimentally~\cite{DL_alfap,DL_eldd,DL_elastic,DL_stot}
\begin{equation}
\alpha_{\pom} = 1.08 + 0.25 t \, .
\label{eq:pomeron_trajectory}
\end{equation}
In Regge theory the energy dependence of total and elastic cross
sections is derived from the analytic structure of the hadronic
amplitudes. In the limit $s \gg -t$, where $s$ is the square of the
center of mass energy of the scattering, the amplitude for elastic
scattering has the form $A(s,t) \prop s^{\alpha_{\pom}(t)}$. The
Pomeron trajectory also provides the leading contribution to the high
energy behavior of the total cross section,
\begin{equation}
\sigma_{{\rm tot}} = s^{-1} {\rm Im} A(s,t=0) \propto s^{\alpha_{\pom}(0)-1}
\label{eq:Regge_xsection}
\end{equation}
The $s$ dependence of hadronic interactions fulfills this behavior
independently of the interacting particles,\cite{DL_stot,DL_stot_rising}
as expected from the universality of the exchanged trajectories.

The deep inelastic charged lepton-proton interactions at low momentum
transfers can be viewed as due to virtual photon proton
interactions, $\gamma^*p$, with the flux of photons originating from
the incoming lepton. The $F_2$ structure function of the proton is
related to the total absorption cross section $\sigma_{{\rm
tot}}(\gamma^*p)$,
\begin{equation}
F_2(x,Q^2) \simeq \frac{Q^2 (1-x)}{4\pi^2\alpha}\sigma_{{\rm
tot}}(\gamma^*p) \, ,
\label{eq:f2}
\end{equation}
where $x$ and $Q^2$ are the Bjorken scaling variable and the negative
of the mass square of the virtual photon respectively. In deep
inelastic scattering (DIS) the kinematical region which corresponds to
the Regge limit is that of low $x$ at fixed $Q^2$ due to the relation,
\begin{equation}
x=\frac{Q^2}{W^2+Q^2-m_p^2} \, ,
\end{equation}
where $W$ denotes the center of mass energy of the $\gamma^*p$ system
and $m_p$ the mass of the proton. For large $Q^2$, DIS can be analyzed
in the framework of perturbative QCD based on the factorization
theorem, the ensuing QCD evolution equations and the quark-parton
model description of the hadrons. The transition between the
perturbative regime and the non-perturbative regime may provide an
understanding of the soft interactions in the language of QCD and
establish whether the Regge model can be justified from first
principles.

The first step in this direction has been made with the measurements
of the electromagnetic proton structure function $F_2$ which cover now
a range in $Q^2$ from $10^{-1}$ to $10^4$ GeV$^2$ and in $x$
from $10^{-6}$ to about 1. The first evidence for soft like
interactions at large $Q^2$ came with the appearance of events with a
large rapidity gap in the hadronic final state in low $x$ interactions
at HERA.\cite{zeus_f2d1,h1_f2d1} These events, when
interpreted in terms of Pomeron exchange, allow a first glance at the
partonic structure of this fundamental ingredient of the Regge
approach. Last but not least the combined data from HERA and LEP start
exploring the structure of the photon in the low $x$ region.

\section{The structure functions of the proton}

\subsection{$F_2$}

\begin{figure}[hbt]
\vspace{-0.5cm}
\center
\psfig{figure=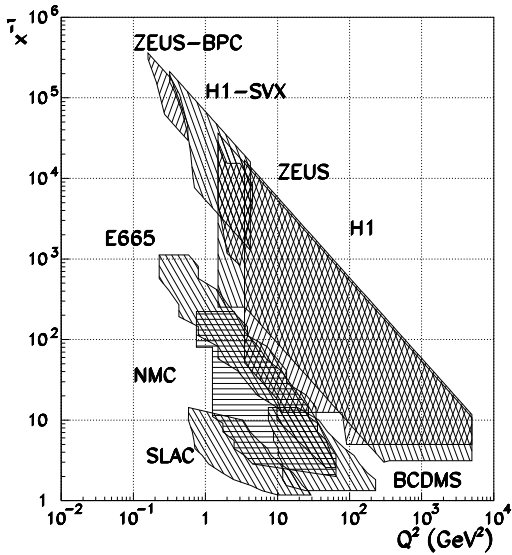,width=7.5cm}
\caption{Phase space coverage of the $F_2$ measurements.}
\label{fig:xq2f2}
\end{figure}

The most precise measurements of the proton structure function with
the widest coverage of the phase space come from electro and
muo-production. At this conference new measurements of the $F_2$
proton structure function have been presented by the
NMC,\cite{nmc_f2last} H1~\cite{h1_f23} and ZEUS~\cite{zeus_f23}
experiments. The existing coverage of the phase space in $1/x$ and
$Q^2$ is summarized in figure~\ref{fig:xq2f2}. Also included are the
previous measurements of the SLAC,\cite{slac_f2}
BCDMS~\cite{bcdms_f21,bcdms_f22} and E665~\cite{e665_f2} experiments.

\begin{figure}[htb]
\vspace{-0.5cm}
\center
\psfig{figure=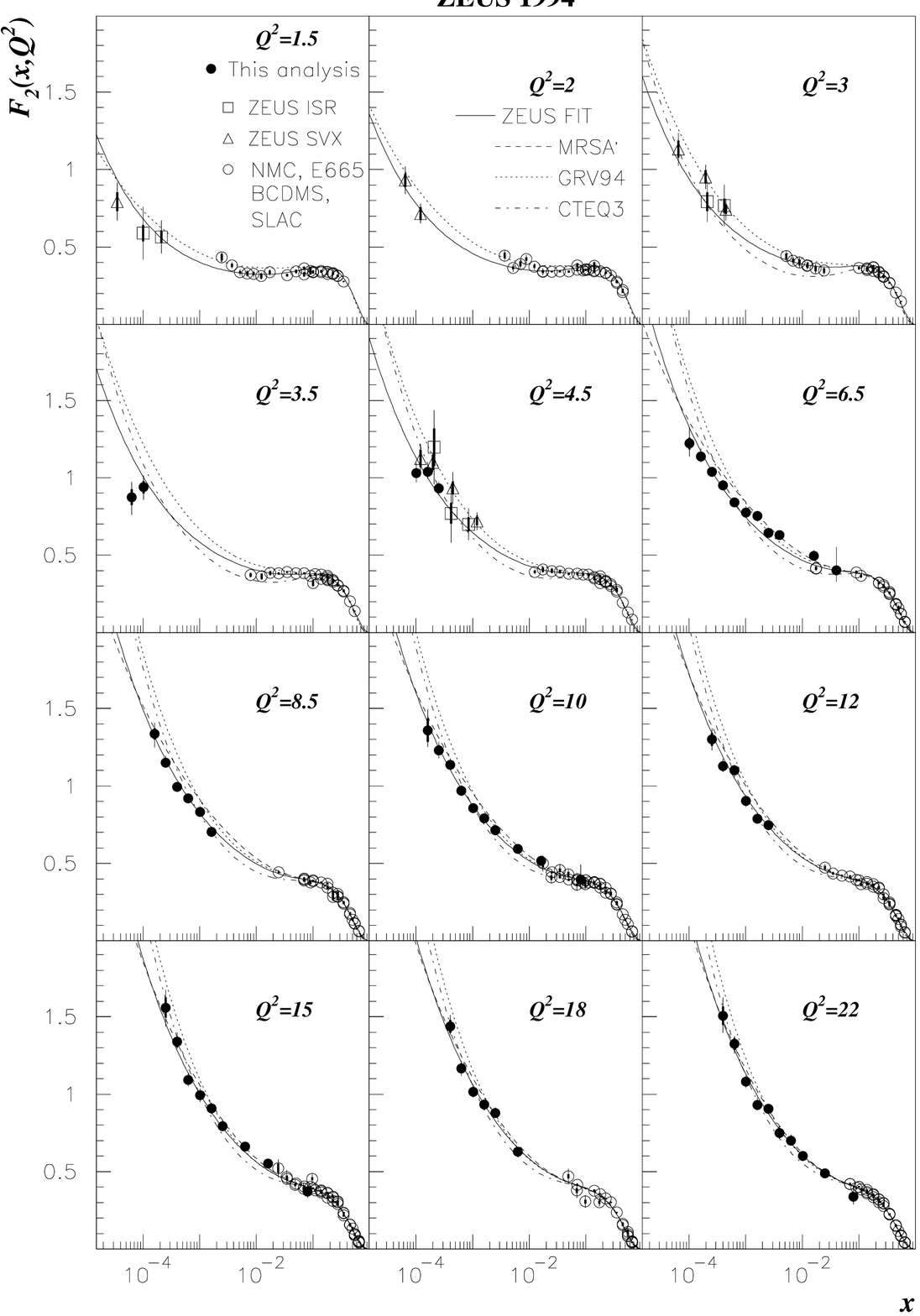,width=8cm}
\vspace{-0.8cm}
\caption\protect{An example of the $x$ dependence of $F_2$ at low $x$ 
as measured by the ZEUS experiment. Also shown are 
selected parameterizations of parton distributions as denoted
on the figure~\cite{mrsap,grv94,cteq3} and the NLO QCD fit
performed by ZEUS.}
\label{fig:f2x}
\end{figure}

One of the striking properties of $F_2$ is that it rises strongly with
decreasing $x$, contrary to expectations based on the Regge approach
and in line with the properties of the QCD evolution
equations.\cite{stirling_rome,sterman_waw} This is shown in
figure~\ref{fig:f2x} where the recent ZEUS
measurements~\cite{zeus_f23} are compared with some of the existing
parameterizations of $F_2$ down to $Q^2=1.5$~GeV$^2$.

\begin{figure}[htb]
\center
%\vspace{-1.cm}
\psfig{figure=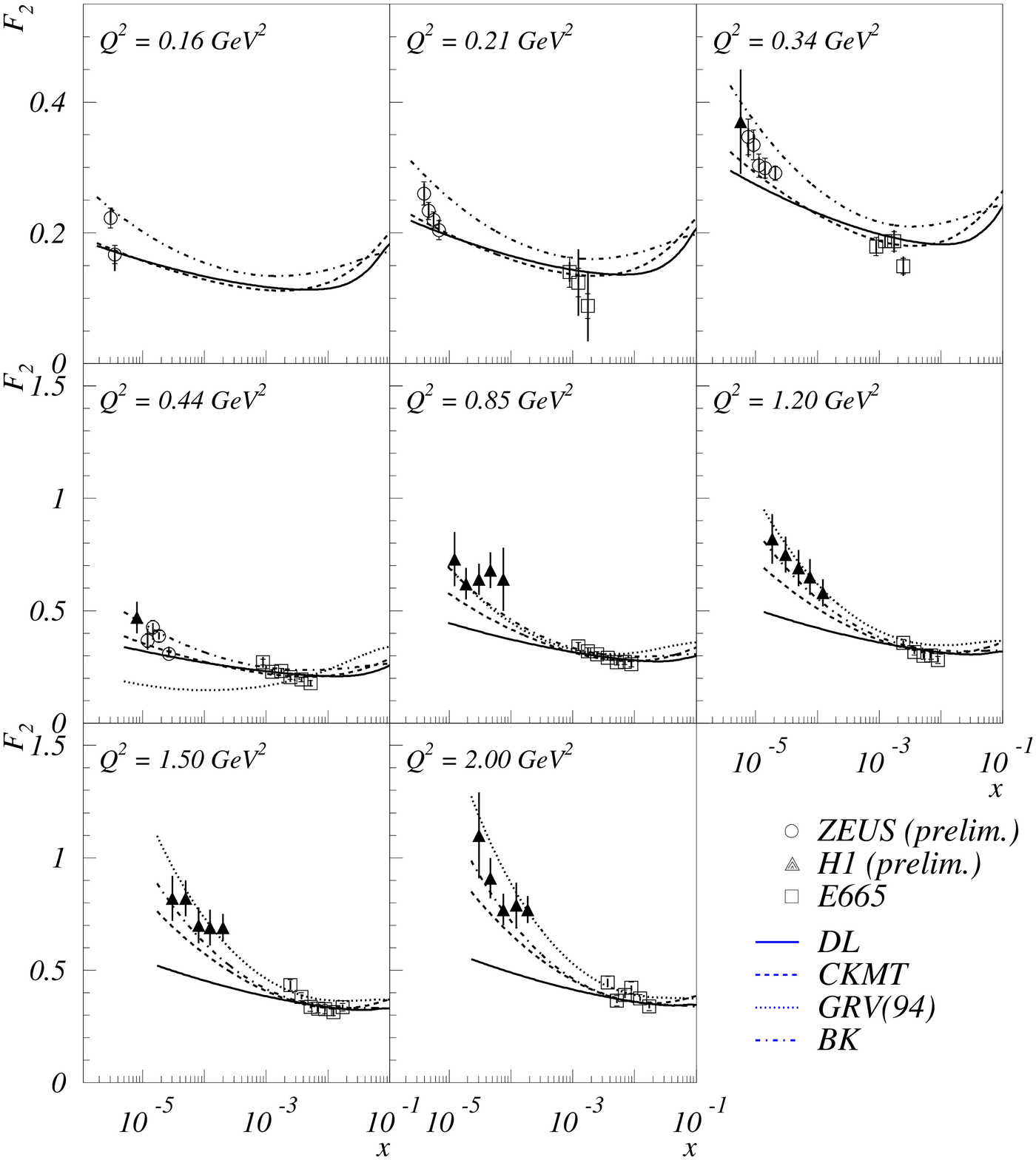,width=7cm}
%\vspace{-0.3cm}
\caption\protect{Preliminary measurements of the proton structure 
function $F_2(x,Q^2)$ at low $Q^2$ by the H1 and ZEUS experiments
together with results from the E665 experiment compared to model
predictions as explained in the text.}
\label{fig:f2_lowq2}
\end{figure}

Since, as has been measured at HERA,\cite{zeus_sigtot,h1_sigtot} the
total photoproduction cross section is known to increase only slowly
with energy, the question which remained opened till this conference
was to establish at which value of $Q^2$ the structure function stops
rising with decreasing $x$. New data were presented by the ZEUS
Collaboration~\cite{zeus_f2_bpc} based on measurements performed with
the Beam Pipe Calorimeter which covers the range $0.16\!<\!Q^2\!<\!0.65
\gev2$. The new measurements of H1~\cite{h1_f2_sv} from a special run 
covering the range $0.35\!<\!Q^2\!<\!3.5 \gev2$ close the remaining
gap in $Q^2$. As seen in figure~\ref{fig:f2_lowq2}, up to $Q^2 \prop
0.85
\gev2$ the data favor parameterizations which are based on the
dominance of soft interactions in this region such as the one of
Donnachie and Landshoff~\cite{dl_f2} or the CKMT~\cite{ckmt_f2} one in
which, due to screening corrections, the effective intercept of the
bare Pomeron is $Q^2$ dependent. For larger $Q^2$ values, $Q^2 \ge 1.2
\gev2$, the GRV~\cite{grv94} and BK~\cite{bk_f2} parameterizations,
perceived in this region as representative of the solution of the NLO
DGLAP QCD evolution equations,\cite{dglap_gl,dglap_d,dglap_ap} give a
very good description of the data.

\begin{figure*}[p]
\psfig{figure=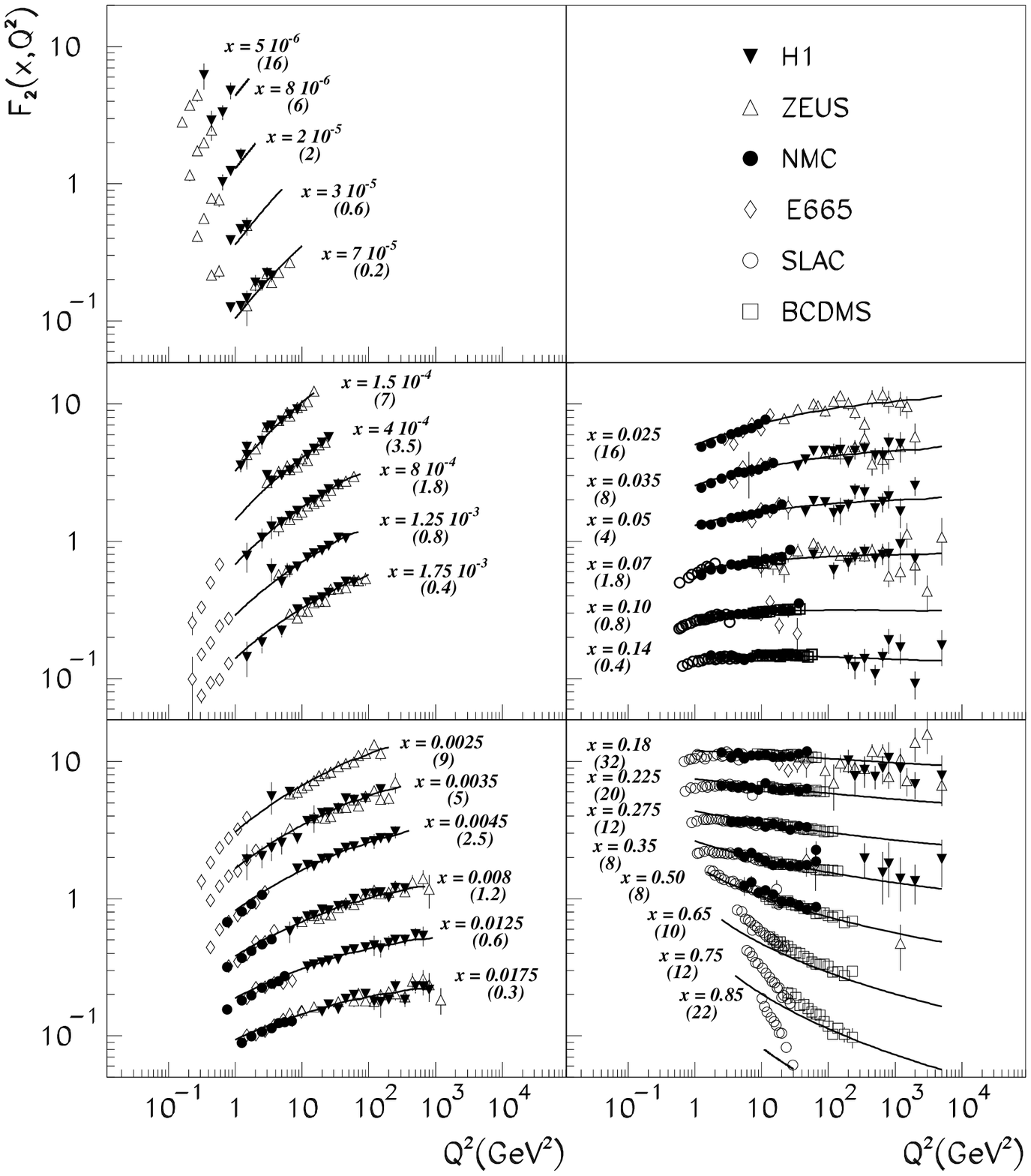,width=16cm}
\caption{Compilation of measurements of $F_2$ as a function 
of $Q^2$ for selected values of $x$ as denoted on the figure. The
numbers in parenthesis are the scaling factors by which the value of
$F_2$ has been multiplied in the plot. The overlayed curves are the
result of an NLO QCD fit performed by the H1 experiment.
\label{fig:f2q2}}
\end{figure*}

The behavior of $F_2$ as a function of $Q^2$ for selected values of
$x$ is presented in figure~\ref{fig:f2q2}. The structure function
$F_2$ rises strongly with $Q^2$ for $x\!<\!0.01$, over three decades in
$Q^2$ in some regions of $x$ where the data are available. With the
new data the gap between the fixed target experiments and the HERA
experiments has been filled and a good agreement in the overlap region
is observed.

\subsection{Structure functions and perturbative QCD}

In perturbative QCD the electromagnetic structure function $F_2$ is
represented as a convolution of the parton distributions and
coefficient functions $C(x,Q^2)$ proportional to the effective
$\gamma^*$-parton couplings,
\bea
\frac{1}{x}F_2(x,Q^2)\!&=&\!\sum_{i=1}^{n_f} e_i^2 C_i(x,Q^2)\otimes 
(q+\bar{q})(x,Q^2) \nonumber \\ 
\!&+&\! C_g(x,Q^2) \otimes g(x,Q^2) \, ,
\label{eq:deff2}
\eea
where $q$ denotes quarks of charge $e_i$ and $g$ gluons, $\otimes$
stands for a convolution integral and $n_f$ is the number of
contributing flavors.  The parton distributions evolve with $Q^2$
following the DGLAP equations,
\be \label{dglap}
{\partial \over \partial \ln Q^2}
 \left(\begin{array}{c} q \\ g \end{array}\right)
 = {\as(Q^2) \over 2 \pi} \left[\begin{array}{cc}
 P_{qq} & P_{qg}  \\
 P_{gq} & P_{gg}
\end{array}\right] \otimes
\left(\begin{array}{c}q \\ g\end{array}\right) \, ,
\ee
where $\as$ denotes the strong coupling constant.  Given a specific
factorization and renormalization scheme, the coefficient functions
$C_i$ and splitting functions $P_{ij}$ are obtained in QCD by
perturbative expansion. In particular
\be
P_{ij}(x,Q^2)= \frac{\as}{2\pi} P_{ij}^{(1)}(x)+
\left(\frac{\as}{2\pi}\right)^2 P_{ij}^{(2)}(x)+ \ldots \, .
\label{eq:pij}
\ee
The truncation after the first two terms in the expansion defines the
conventional NLO DGLAP evolution. This evolution assumes that the
dominant contribution to the DIS cross section comes from parton
cascades in which partons from subsequent emissions are strongly
ordered in transverse momenta $k_t$, the largest corresponding to the
parton interacting with the probe.

At low $x$ higher-loop contributions to the splitting functions are
enhanced,
\be
P_{ij}^{(n)} \sim \frac{1}{x} \ln^{(n-1)} x \, .
\ee
The presence of these terms may spoil the convergence
of~(\ref{eq:pij}). The evolution equation which allows the resummation
of leading $(\as \ln x)^n$ contributions and which describes parton
radiation without $k_t$ ordering is known under the name
BFKL.\cite{bfkl_1,bfkl_2,bfkl_3}  In the parton cascade picture this
contribution corresponds roughly to cascades which develop from a
large $k_t
\sim Q$ parton emitted at large $x$ and the rest of the evolution
takes place only in $x$.  The two approaches, the DGLAP and the BFKL
one are embodied in the CCFM equation,\cite{ccfm_1,ccfm_2,ccfm_3}
based on the $k_t$ factorization and angular ordering.

The solutions of the DGLAP equations and of the BFKL equation, in
the limit of very low $x$, where the dominant contribution to the
cross section is driven by gluon radiation, predict a rise of $F_2$
with decreasing $x$, 
\be
F_2^{DLL}(x,Q^2) \sim \exp\left(2 \sqrt{\frac{C_A\as}{\pi} 
\ln \frac{1}{x}
\ln \frac{Q^2}{Q^2_0}}\right) \label{eq:dlla} \, ,
\ee
\be
F_2^{BFKL}(x,Q^2)  \sim  \sqrt{\frac{Q^2}{Q^2_0}} 
x^{-\frac{4C_A\as}{\pi}\ln 2}
\label{eq:bfkl}
\ee
where the superscript $DLL$ stands for the double leading logarithmic
approximation used in solving the DGLAP equation and $Q_0^2$ denotes
the starting scale of the evolution. In general the BFKL equation
predicts a faster increase of $F_2$ with decreasing $x$ and stronger
scaling violations in $Q^2$ as compared to the DGLAP evolution.
However the solution~(\ref{eq:bfkl}) is derived assuming a constant
$\as$ and higher order corrections are expected to tame the rise with
$1/x$.\cite{adm_rome} In the BFKL approach the concept of a QCD
Pomeron arises naturally.

The confrontation of the various approaches with the data requires
assumptions on the input parton distributions which cannot be derived
from first principles. The solution of the NLO DGLAP evolution
equations with input parton distributions fitted to the data has been
widely explored~\cite{grv94,cteq3,MRS,mrs96,GRV,CTEQ} and gives a
very good description of the data, down to amazingly low values of
$Q^2$. An example is shown in figure~\ref{fig:f2q2} where the NLO
parameterization of $F_2$ obtained by H1~\cite{h1_f23} is
presented. The data were fitted for $Q^2>7 \gev2$ and evolved
backwards down to $Q^2=1 \gev2$.

\begin{figure}[htb]
\center
\psfig{figure=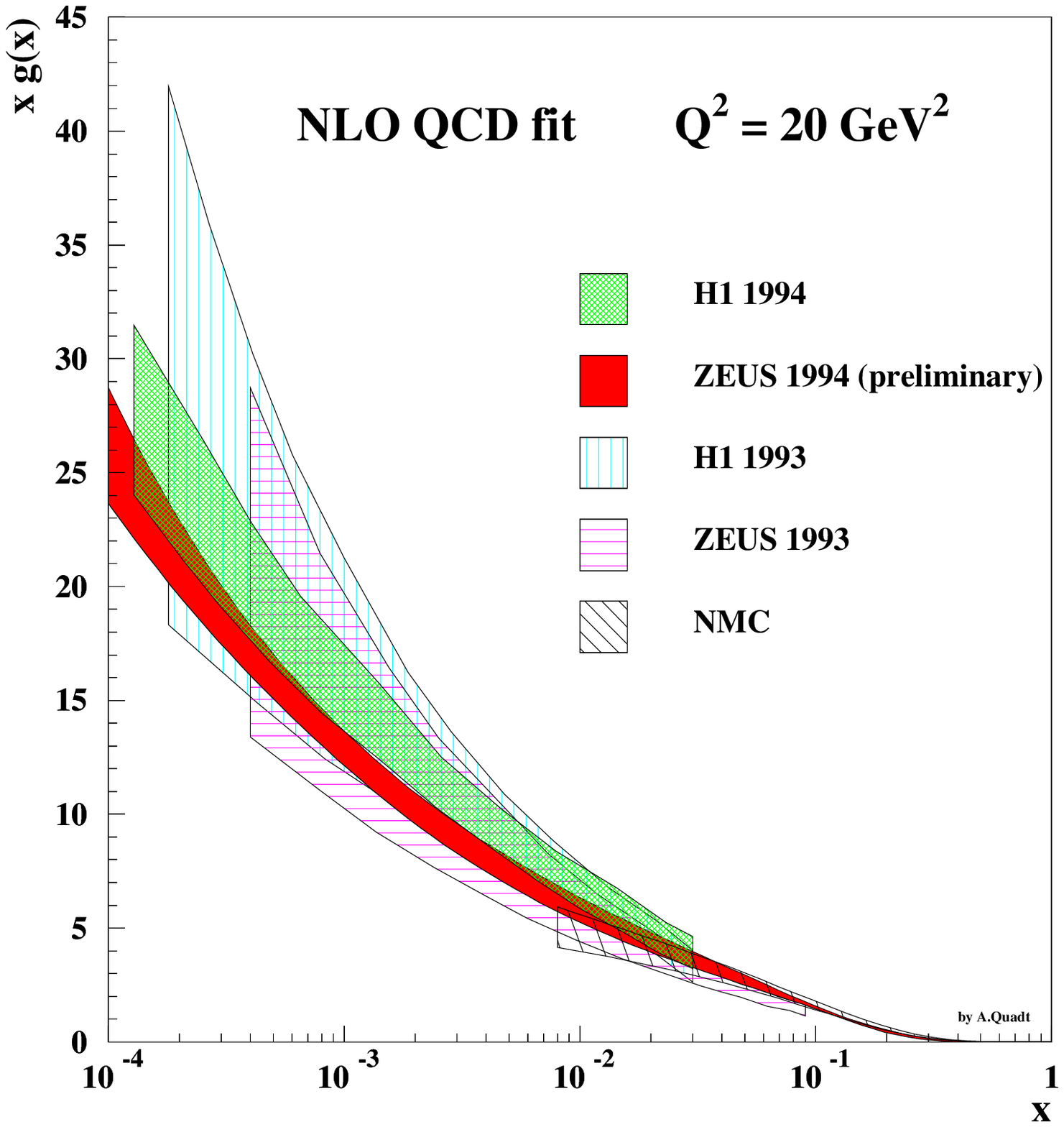,width=7cm}
\vspace{-0.3cm}
\caption\protect{Gluon distribution in the proton as determined 
from NLO QCD fits to the $F_2$ data.
}
\label{fig:gluons}
\end{figure}

A byproduct of the NLO DGLAP fits to the data is the NLO gluon
distribution in the proton. If the value of $\as$ is known the gluon
distribution at low $x$ is constrained by the logarithmic slope of
$F_2$ in $Q^2$. The gluon distributions obtained at $Q^2=20 \gev2$ by
the respective experimental groups, which can handle best the
correlated systematic errors, are presented in
figure~\ref{fig:gluons}. Also shown for comparison are the results
obtained earlier with smaller statistics. The improvement is evident.
One may be led to conclude that, since the data are perfectly
described by the NLO DGLAP, there is no evidence for new QCD dynamics
at low $x$. However one should keep in mind that a set of equations
based only on $F_2$ measurements is under-constrained. This is
because, while $F_2$ is directly proportional to the quark
distributions (see equation~(\ref{eq:deff2})),
\be
{\partial F_2 \over \partial \ln Q^2} \propto P_{gg} \otimes g \, ,
\ee
and the gluon distribution is reliable only to the extent to which
$P_{gg}$ is known. Ways of constraining the system are to measure
either the longitudinal structure function $F_L$ and/or the charm
contribution to $F_2$~.\cite{catani_2,catani_3,roberts}

\subsection{$F_L$}

The lack of knowledge of $F_L$ limits the experimental accuracy with
which $F_2$ can be determined. The relation between the measured
electromagnetic differential DIS cross section and the $F_2$ and $F_L$
structure functions is the following:
\be
\frac{d^2\sigma}{dxdQ^2}=\frac{2\pi\alpha^2}{xQ^4}\left[ \left( 1+(1-y)^2
\right) F_2 - y^2F_L \right] \, .
\label{eq:xsec_dis}
\ee
At low $x$, that is high $y$, the contribution of $F_L$
is non-negligible and cannot be ignored.  The usual procedure adopted
experimentally is to use, in the region of moderate $x$, the
parameterization of the ratio $R \approx F_L/(F_2-F_L)$ based on the
dedicated SLAC measurements~\cite{slac_r90} and, at low $x$, the QCD
expectations based on existing parton distributions.
\begin{figure}[hbt]
\center
\vspace{-0.5cm}
\psfig{figure=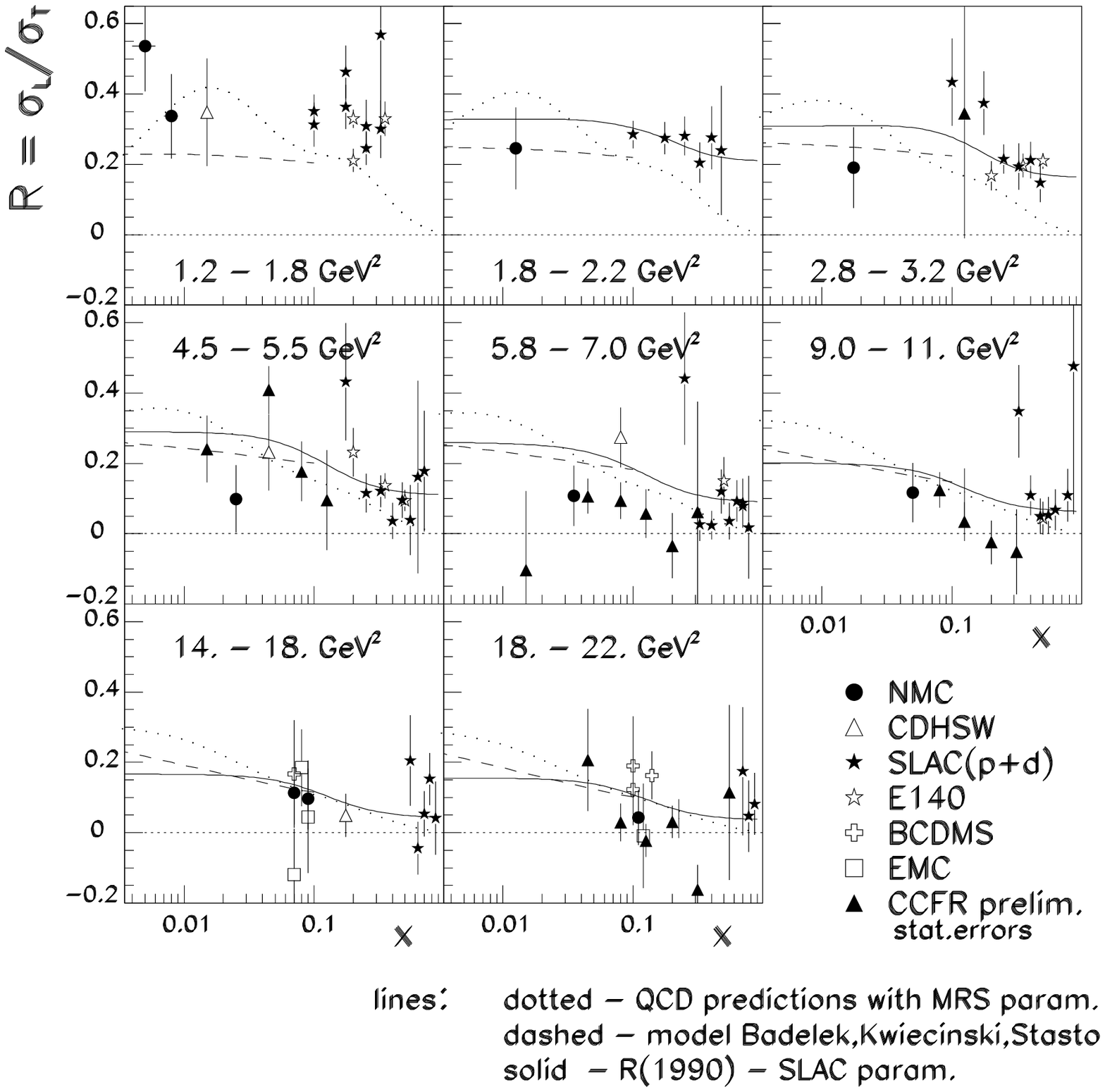,width=7cm}
\caption\protect{Measurements of $R$ as a function of $x$ in bins of 
 $Q^2$, compared to the SLAC (full line), QCD NLO MRS(R1) (dotted line) and
 Bade\l ek et al. (dashed line) parameterizations.  }
\label{fig:fl}
\end{figure}

The experimental determination of
$R=\sigma_L(\gamma^*p)/\sigma_T(\gamma^*p)$, where $L,\,T$ stand for
the longitudinal and transverse polarization of the virtual photon,
requires the measurement of the differential cross
section~(\ref{eq:xsec_dis}) at different center of mass energies and
fixed $x,~Q^2$. New measurements of $R$ have been presented at this
conference by the NMC Collaboration~\cite{nmc_f2last} which has used
muo-production data taken at four different muon beam momenta ranging
from 90 to 280~GeV. Preliminary results have been also presented by
the CCFR Collaboration~\cite{ccfr_fnu,ccfr_fl} based on
$\stackrel{{\tiny (-)}}{\nu}\mathrm{Fe}$ in the neutrino energy range
from 10 to 350~GeV. The new measurements extend the range of $x$ down
to $x \sim 10^{-3}$ at the lowest $Q^2 \sim 1.5\gev2$. The compilation
of all existing data is presented in figure~\ref{fig:fl}.
The measurements are compared to three different parameterizations,
the SLAC one,\cite{slac_r90} the new parameterization of Bade\l ek at
al.\cite{bk_fl} presented at this conference and to the QCD
expectations based on the new MRS(R1) parameterization of parton
distributions.\cite{mrs96} As can been seen from this comparison all
of them reproduce the data fairly well, although they may differ up
to $50 \%$ at low $x$.

\begin{figure}[htb]
\center
\psfig{figure=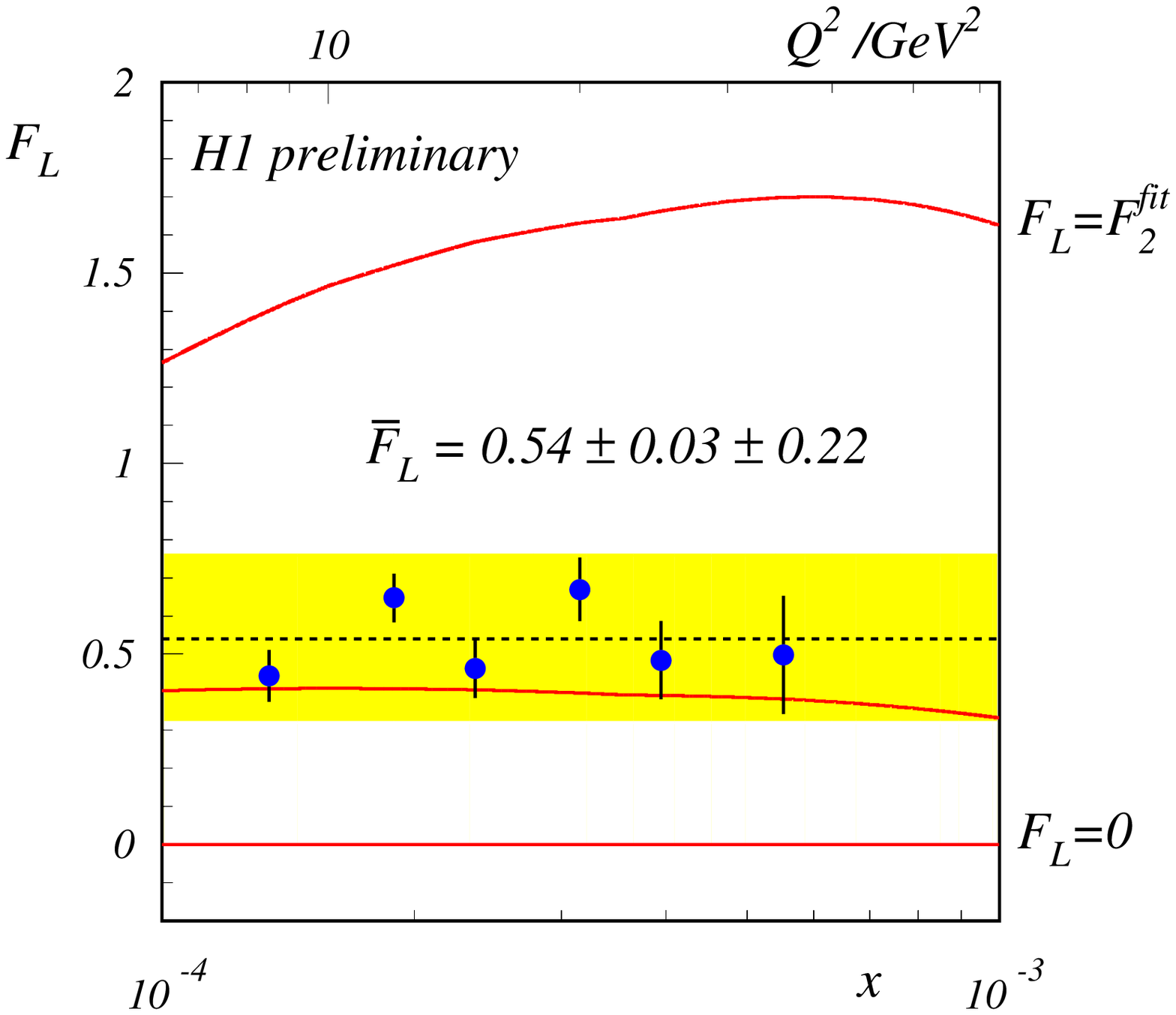,width=7cm}
\vspace{-0.4cm}
\caption\protect{$F_L$ as a function of $x$ (and $Q^2$) as determined 
by the H1 experiment. The full line closest to the data corresponds to
NLO QCD expectations.  }
\label{fig:h1_fl}
\end{figure}

In order to constrain the value of $F_L$ in the low $x$ regime
explored at HERA, the H1 Collaboration determined $F_L$ assuming that
the NLO DGLAP evolution holds.\cite{h1_f2_sv} The parameterization of
$F_2$ was obtained using only the data for $y\!<\!0.35$, where the
contribution of $F_L$ is small (see formula~(\ref{eq:xsec_dis})). The
difference between the value of $F_2$ expected from this
parameterization at larger $y$ and the one determined from the cross
section assuming $F_L=0$ is then used to determine the required $F_L$.
The results of this procedure are shown in
figure~\ref{fig:h1_fl}. Also shown are the expectations for $F_L$
obtained from the same QCD fit and assuming that $F_L=F_2$. Within
errors there is a consistency between the values obtained from the
procedure above and the QCD fit, again pointing to the fact that the
NLO QCD evolution is describing the data very well.

\subsection{$F_{2}^{c\bar{c}}$}

The measurement of charm production is of utmost
importance. The perturbative QCD calculations involving heavy quarks
are most reliable due to the large masses involved. It may turn out
that charm production cross sections will provide the best constraint
on the gluon distribution in the proton.\cite{durham_charm}
Experimentally though, large integrated luminosities are needed
because of the low tagging efficiency of charm in the final state.

\begin{figure}[htb]
\center
\psfig{figure=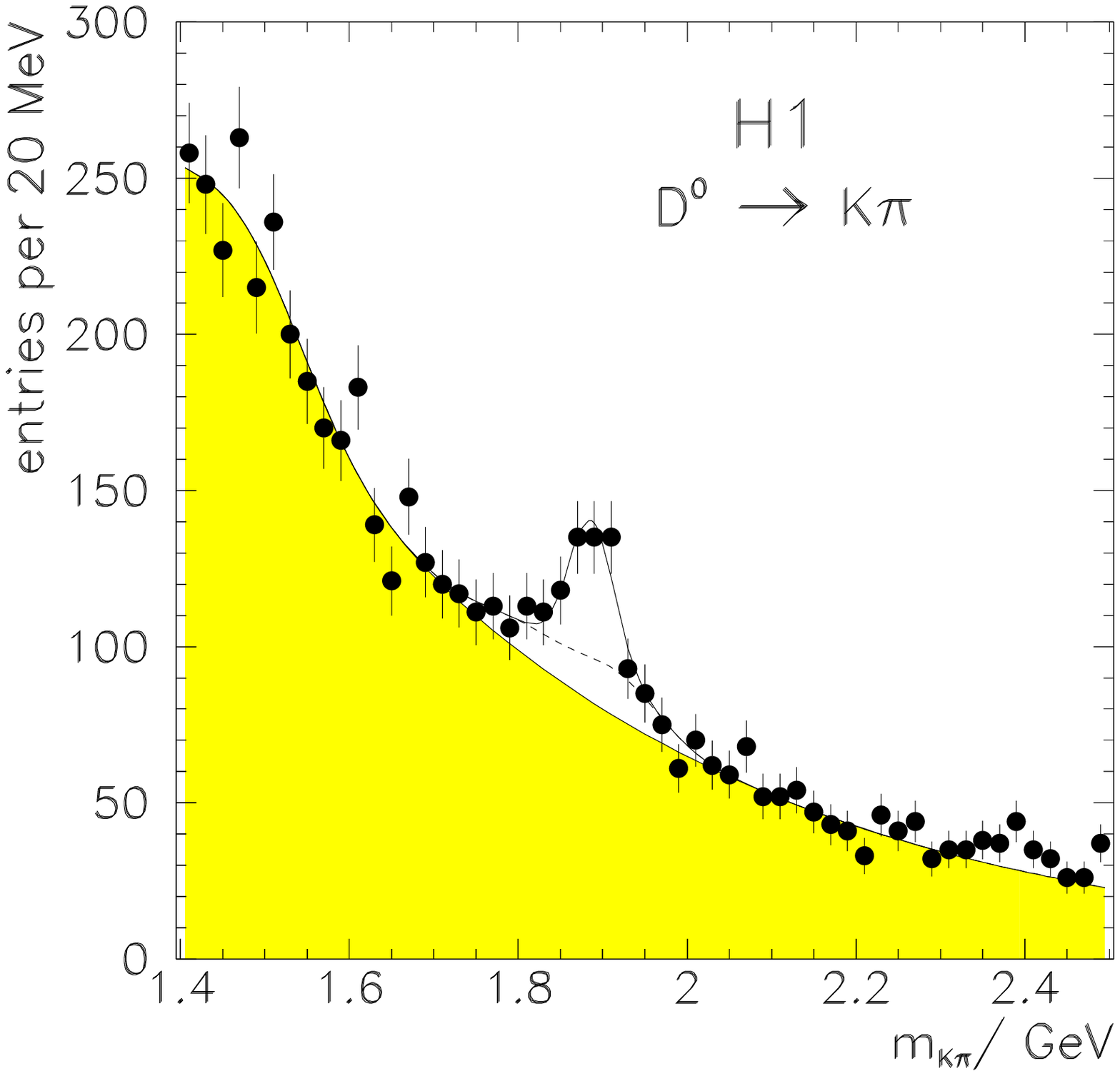,width=8cm}
\vspace{-1.cm}
\caption\protect{The invariant mass spectrum of $K\pi$ as seen in the 
H1 detector.}
\label{fig:h1_d0}
\end{figure}

The first glance at the charm production properties in DIS have been
presented by the H1~\cite{h1_f2_sv} and ZEUS~\cite{zeus_charm}
experiments, based on statistics of the order of 100 events or
so. Experimentally charm production was tagged by reconstructing the
$D^0$ (H1) or the $D^*$ (H1 and ZEUS) mesons and their charge
conjugates. The signal of $D^0 \rightarrow K \pi$ decay as seen in the
H1 detector is presented in figure~\ref{fig:h1_d0}.
\begin{figure}[htb]
\center
\psfig{figure=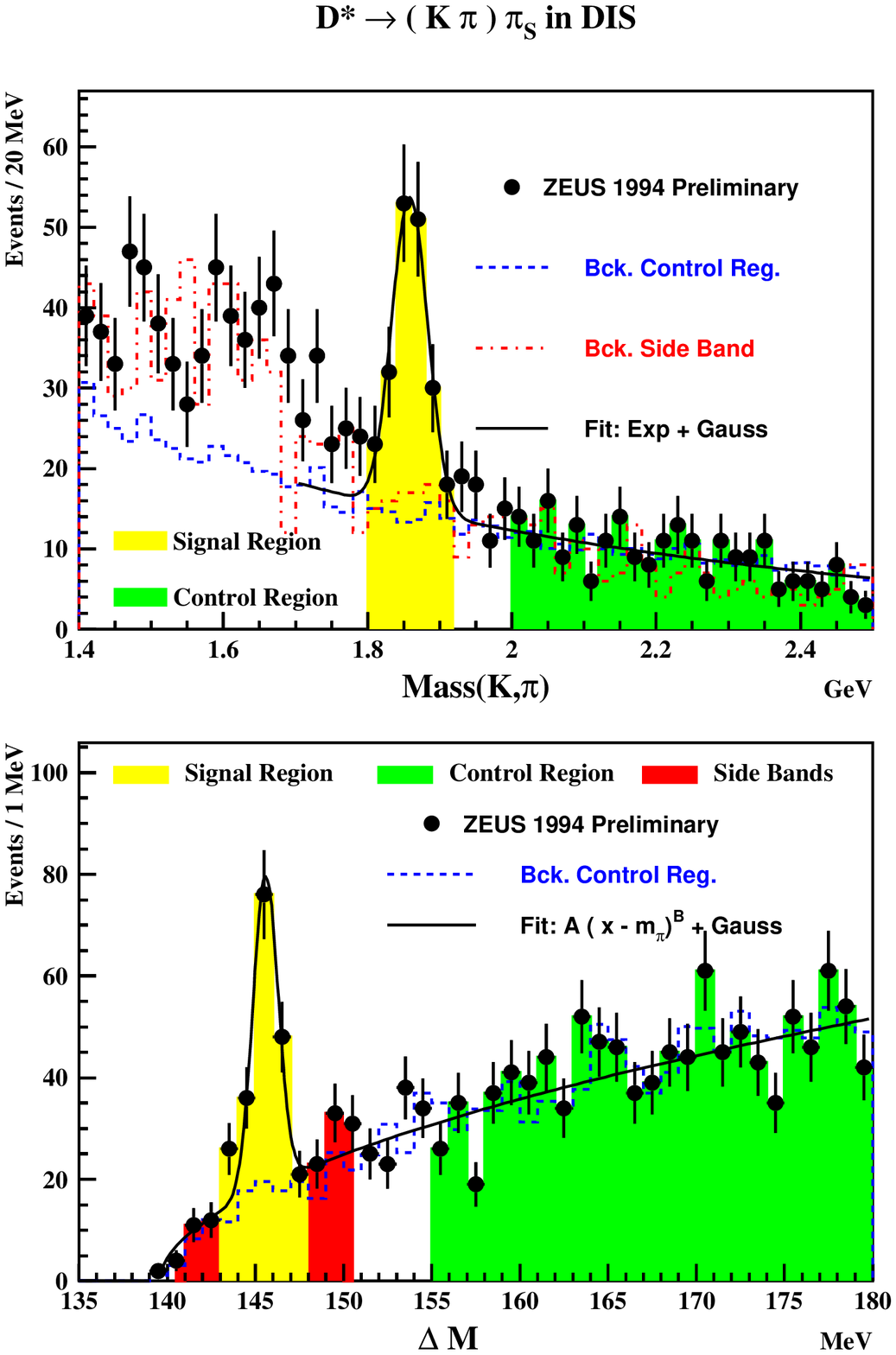,width=8cm,%
bbllx=70pt,bblly=110pt,bburx=520pt,bbury=400pt,clip=}
\vspace{-0.1cm}
\caption\protect{The $\Delta M =m (K\pi\pi_s)-m(K\pi)$ distribution for 
$m(K\pi)$ in the $D^0$ signal region, where $\pi_s$ denotes the
softest pion.}
\label{fig:zeus_dstar}
\end{figure}
The $D^*$ decays
are identified by the decay chain $D^* \rightarrow D^0 \pi \rightarrow
K \pi \pi$, taking advantage of the tight kinematic constraint imposed
by the small mass difference $\Delta M = m_{D^*}-m_{D^0} = 145.5 \pm
0.15$~MeV.  As an example the signal observed in the $\Delta M$
distribution in the ZEUS analysis is shown in
figure~\ref{fig:zeus_dstar}.  

The dominant mechanism of charm production in DIS at HERA energies is
the boson-gluon fusion process. Indeed the AROMA~\cite{aroma} MC based
on the boson gluon fusion mechanism is found to reproduce properly the
shapes of the transverse momentum and pseudorapidity of the $D$ mesons
as well as the overall $W$ and $Q^2$ dependence. An upper limit of
$5\%$ for a possible contribution of the charm sea has been estimated
by the H1 experiment.  

The ZEUS experiment finds that also the absolute value of the measured
cross section is well reproduced by the boson gluon fusion mechanism
calculated to LO, if the appropriate GRV gluon
distribution~\cite{grv94} is used in the simulation and if the mass of
the charm quark is kept within the limits $1.35 \le m_c
\le 1.5$~GeV. 
\begin{figure}[hbt]
\center
\vspace{-0.3cm}
\psfig{figure=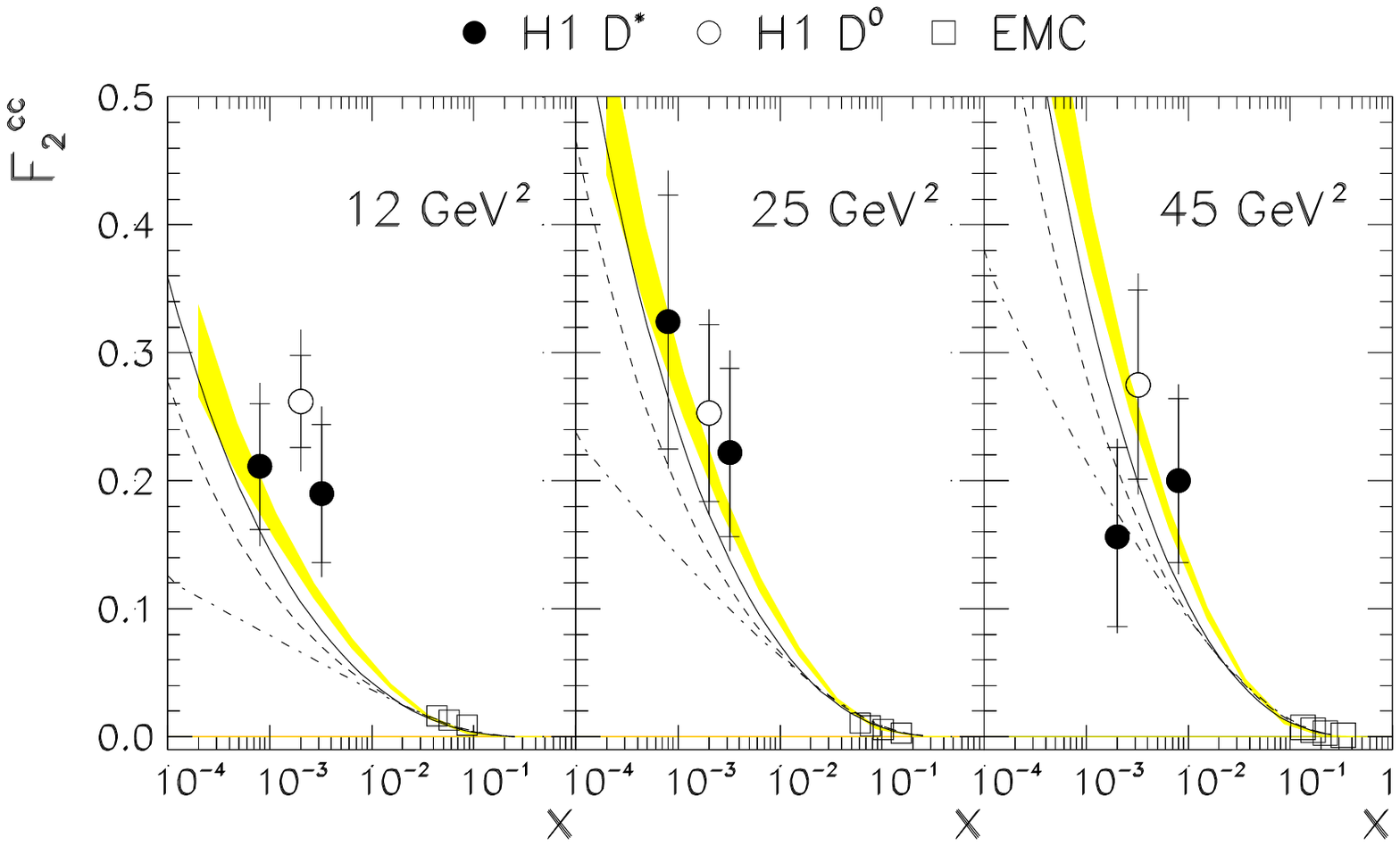,width=8cm}
\vspace{-0.7cm}
\caption\protect{$F_2^{c\bar{c}}$ as measured by the H1 experiment. 
The curves correspond to various QCD parton parameterizations. The
shaded area is the expectation based on the H1 NLO QCD fit.}
\label{fig:h1_f2charm}
\end{figure}

The charm production cross section measured by the H1
experiment is found to be larger than expected from NLO calculations
for any known gluon distributions. This can partly be seen in
figure~\ref{fig:h1_f2charm} where the $F_2^{c\bar{c}}$ as determined
by H1 is compared to NLO QCD expectations.

Comparison of the H1 results with the EMC~\cite{EMC_charm} data
reveals a steep rise of $F_2^{c\bar{c}}$ from $x \sim 0.1$ to $
10^{-3}\!<\!x\!<\!10^{-2}$. Averaged over the measured kinematic range of H1 a
ratio $\langle F_2^{c\bar{c}}/F_2\rangle = 0.237 \pm
0.021^{+0.043}_{-0.039}$ is obtained, one order of magnitude larger
than at larger $x$.

\subsection{$\alpha_s$ from DIS}

An important ingredient of the perturbative QCD analysis of the $F_2$
structure function is the value of $\alpha_s(M_Z^2)$ assumed in the
fits. The CCFR Collaboration has presented preliminary results from
their study of $\stackrel{{\tiny(-)}}{\nu}\mathrm{Fe}$
interactions.\cite{ccfr_fnu} The evolution of the $xF_3$ structure
function allows a direct determination of $\alpha_s$. The preliminary
value presented at this conference is $\alpha_s(M_Z^2) = 0.118 \pm
0.007 $ to be compared with $0.113 \pm 0.005$ previously
determined from a combined fit to the SLAC/BCDMS $F_2$
evolution~\cite{vm_f2fit} and $0.122 \pm 0.004$ from the global LEP
fit to the electroweak data.\cite{busenitz} The good news is that now
the value of $\alpha_s$ determined from the DIS and the LEP data are
in good agreement.

\subsection{Conclusions}

The structure function of the proton $F_2$ is found to be rising with
decreasing $x$ down to $Q^2$ as low as $1 \gev2$. At moderate $Q^2
\sim 10 \gev2$, the rise can be quantified by $F_2 \sim x^{-0.2 \div
-0.3}$ and it becomes faster at higher $Q^2$. The observed trend is
very well described by the conventional DGLAP evolution in
NLO. However this observation does not disprove the existence of new
dynamics at low $x$. This is because the system is under-constrained
and a combination of unknown parton distributions with different
splitting functions can mimic the observed behavior of
$F_2(x,Q^2)$. To test this point measurements of the longitudinal
structure function $F_L$ and/or the charm contribution to $F_2$ are
essential. The existing measurements are far too imprecise.  Some
indications that the $k_t$ ordering implied by the DGLAP evolution may
not be the full story comes from studies of the hadronic final
states.\cite{h1_hadpt} Another reason to look beyond the standard
evolution is that $F_2(x,Q^2)$ cannot rise indefinitely with
decreasing $x$ without violating the unitarity bound.

\section{Diffractive Hard Scattering}

\begin{figure}[htb]
\center
\epsfig{figure=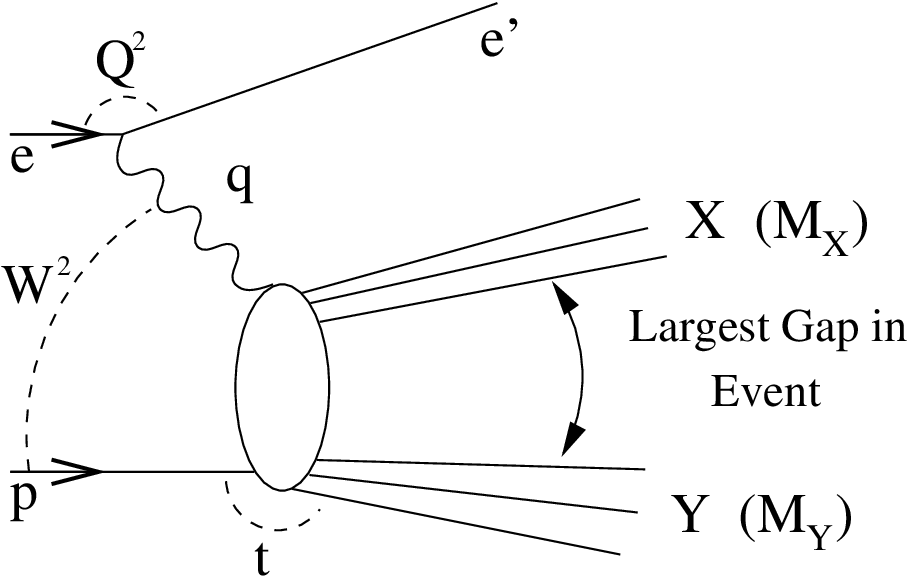,height=7cm,angle=-90.}
\caption\protect{Schematic representation of a DIS event with a 
rapidity gap.}
\label{fig:diag_lrg}
\end{figure}

\subsection{Inclusive DIS diffraction}

One of the surprises which came along with the first results from HERA
was the observation of DIS events with a large rapidity gap in the
hadronic final state, located between the photon and the proton
fragmentation regions~\cite{zeus_f2d1,h1_f2d1} as depicted in
figure~\ref{fig:diag_lrg}. In the fragmentation picture driven by
parton radiation large rapidity gaps are exponentially
suppressed.\cite{basics_pqcd} The observed fraction of events with
large rapidity gaps is of the order of $10 \%$ fairly independent of
$W$ and $Q^2$.  This property is typical of diffractive scattering and
invokes the notion of the Pomeron. The fact that the diffractive
exchange can be probed with high $Q^2$ virtual photons means that its
structure can be studied much the same way as the partonic structure
of the proton.
\begin{figure*}[htb]
\psfig{figure=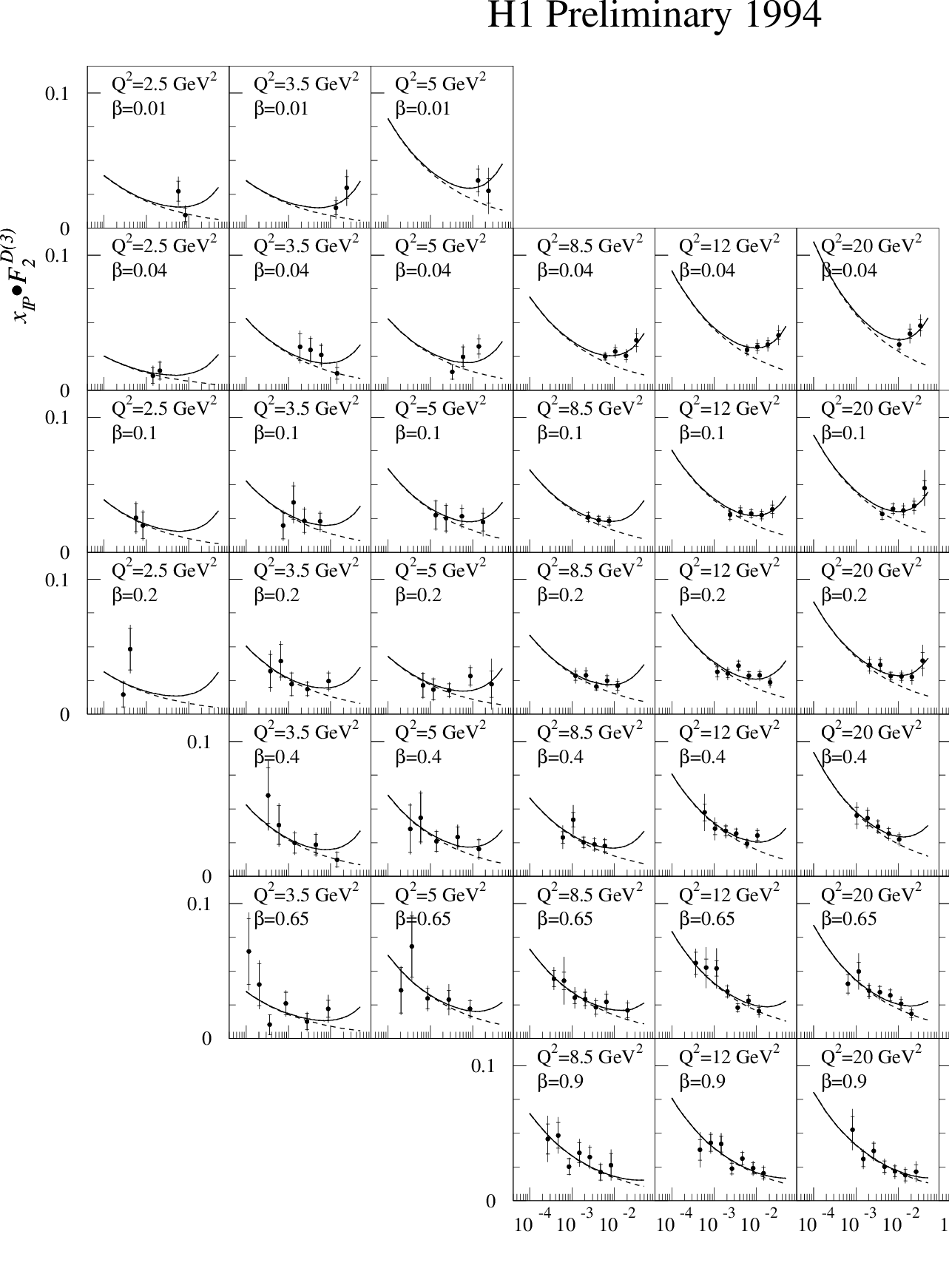,height=13cm,width=11cm}
\caption{$x_{\pom}F_2^{D(3)}$ as a function of $x_{\pom}$ in bins of $\beta$
and $Q^2$ as denoted on the figure. The full line is the result of the 
fit of an incoherent sum of the $\pom$ and $f$ meson trajectories. The
dashed line corresponds to the $\pom$ contribution.
\label{fig:h1_f2d3}}
\end{figure*}

Diffractive events in DIS at low $x$ were
expected~\cite{bj_1,dl_dis_pom,IS_pom,nikzak_pom} long before they
were observed. There are various ways of understanding it. The most
compelling way is to think of the $\gamma^*p$ interaction in the rest
frame of the proton. In this frame the life time of a $q \bar{q}$
fluctuation of the virtual photon can be estimated to be~\cite{halfms}
\be
\tau_{q \bar{q}} \simeq \frac{1}{2m_px} \, ,
\ee
where $m_p$ denotes the mass of the target proton. At $x=10^{-4}$,
$c\tau_{q \bar{q}} \simeq 10^3~\mathrm{fm}$ is much larger than the
typical size of a hadronic target. The photon may thus fluctuate into
a $q \bar{q}$ pair long before arrival on target. The preferred
configuration is the one in which the quarks have a relatively small
transverse momentum, form a large size object and the virtual photon
turns into a hadronic probe. It is thus to be expected that the
$\gamma^*p$ interactions will be similar to hadron-hadron
interactions. Of course QCD introduces corrections to the above
picture, such as scaling violation or production of large $p_T$ jets,
all of which can be accommodated in the laboratory frame
description.\cite{hera_ws96,buchmuller} In this picture diffractive
scattering emerges naturally, as a consequence of the hadronic-like
nature of the virtual photon.

Assuming the validity of Regge factorization for $\gamma^*p$
diffractive interactions (while ignoring the sub-leading Regge
trajectories) we expect,
\be
\frac{d^2 \sigma(\gamma^*p)}{dt dx_{\pom}} = F_{\pom/p}(t,x_{\pom}) 
\sigma(\gamma^* \pom) \, ,
\label{eq:pomflux}
\ee
where $ F_{\pom/p}(t,x_\pom)$ denotes the flux of the Pomeron in the
proton, $x_\pom$ is the fraction of the proton momentum carried by the
Pomeron and $t$ is the square of the four-momentum transfer at the
proton vertex. We may define the DIS differential cross section (and
the appropriate structure function) for producing a diffractive
hadronic state $X$ of mass $M_X$ in the reaction $ep \rightarrow epX$,
\be
\frac{d^3\sigma^D}{dx_{\pom} d\beta dQ^2}=\frac{2\pi \alpha^2}{\beta Q^4}
\left[ 1+(1-y)^2 \right]  F_2^{D(3)}(\beta,Q^2,x_{\pom}) \, ,
\ee
where $\beta$ is defined as
\be
\beta = \frac{x}{x_{\pom}} \simeq \frac{Q^2}{M_X^2+Q^2} \, ,
\ee
and an implicit integration over $t$ is assumed in accordance with the
experimental conditions. Should the Regge type factorization hold in
the DIS regime, one expects
\be
F_2^{D(3)}(\beta,Q^2,x_{\pom}) = \frac{1}{x_{\pom}^n} F_2^{\pom}(\beta,Q^2) \, ,
\label{eq:factorization}
\ee
with the structure function of the Pomeron, $F_2^{\pom}$, defined by
this equation. The first term stands for the flux of the Pomeron with
$n=1-2 \alpha_{\pom}(0) + \delta_t$, independent of $\beta$ and $Q^2$
($\delta_t \simeq 0.03$ is a correction due to the effect of $\alpha'$
in the integration over $t$). Experimentally $F_2^{D(3)}$ can be
defined either at the hadronic level, by the presence of a large
rapidity gap between $X$ and the final state of the proton $Y$ (see
figure~\ref{fig:diag_lrg}) which for most analyzes remains undetected
or by tagging a fast proton in the forward spectrometer.  

\begin{figure}[htb]
\center
\epsfig{figure=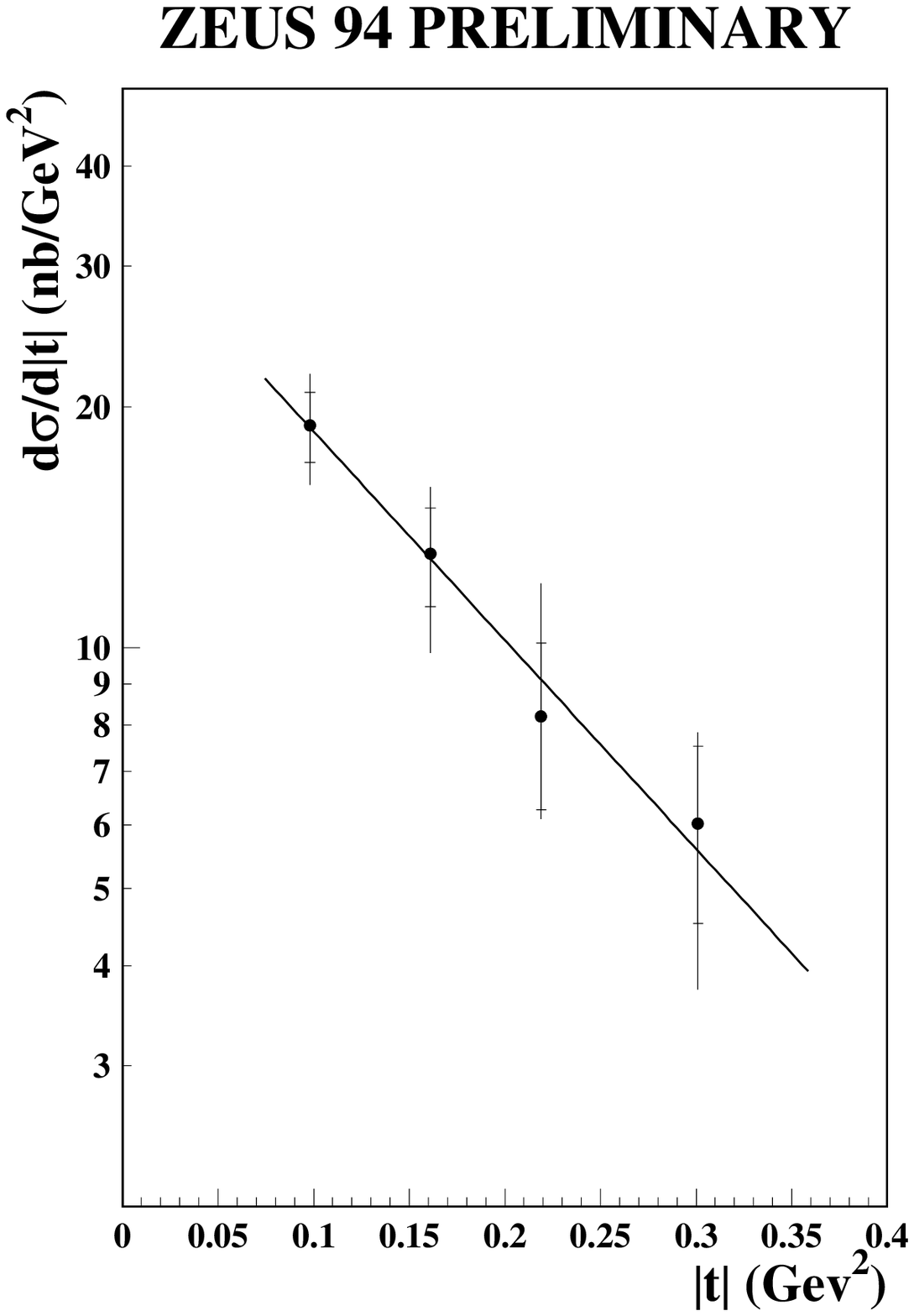,height=7cm}
\caption\protect{The $t$ distribution of diffractive DIS as measured
 with the LPS.}
\label{fig:lps_t}
\end{figure}

Tests of factorization have been conducted by the H1
Collaboration~\cite{h1_f2d3} in the range $2.5\!<\!Q^2\!<\!65 \gev2$,
$0.01\!<\!\beta\!<\!0.9$ and $10^{-4}\!<\!x_{\pom}\!<\!5\!\cdot\!
10^{-2}$. As can be seen in figure~\ref{fig:h1_f2d3} a change of the
$x_{\pom}F_2^{D(3)}$ dependence on $x_{\pom}$ as a function of $\beta$
is observed. However if a fit to $F_2^{D(3)}$ is performed assuming a
contribution of both the Pomeron and the $f$ meson trajectories the
results are consistent with the intercepts of the two trajectories
being independent of $\beta$ and $Q^2$ as expected for Regge
factorization. The intercept of the $f$ trajectory is compatible with
the expectations ($\alpha_f\simeq 0.6$) while that of the Pomeron,
\[\alpha_{\pom}(0) = 1.18 \pm 0.02 (\mathrm stat) \pm 0.07 (\mathrm syst) 
\, , \] 
is within errors marginally higher than would be expected for a soft
Pomeron.~\footnote{The value of $\alpha_{\pom}$ measured in the energy
dependence of the DIS diffractive cross section by the ZEUS
Collaboration~\cite{zeus_f2d3,zeus_lps} has been found substantially
larger than the value from soft interactions. However recently a
technical mistake has been found in the generation of the Monte Carlo
data used for the acceptance correction and resolution unfolding. This
mistake led to the mishandling of QED radiative corrections. Its
effect is to change the cross sections by typically one systematic
error. The ZEUS collaboration thus has retracted their results until
further analysis is completed.}  A similar intercept,
\[\alpha_{\pom}(0) = 1.17 \pm 0.04 (\mathrm stat) \pm 0.08 (\mathrm syst)
\, , \]
has been obtained by the ZEUS Collaboration in a measurement performed
with the Leading Proton Spectrometer~\cite{zeus_lps} (LPS) at an
average $Q^2=12 \gev2$ for the range $0.006\!<\!\beta\!<\!0.5$ and
$4\!\cdot\!10^{-4}\!<\! x_{\pom}\!<\! 3 \!\cdot\!10^{-2}$. In this
measurement the scattered proton, with a fraction of the initial
proton momentum larger than 0.93, is detected in the LPS. The
advantage of detectors such as the LPS is that the $t$ distribution of
the diffractive events can be studied. The $|t|$ distribution measured
in the range $0.07\!<\!|t|\!<\!0.35
\gev2$, corrected for detector acceptance and resolution, is presented
in figure~\ref{fig:lps_t}. Assuming an exponential fall-off in $|t|$,
the fitted slope is determined to be $b = 5.9 \pm
1.3^{+1.1}_{-0.7}$~GeV$^{-2}$. This value is within errors compatible
with expectations based on the soft Pomeron phenomenology.

\begin{figure}[htb]
\center
\epsfig{figure=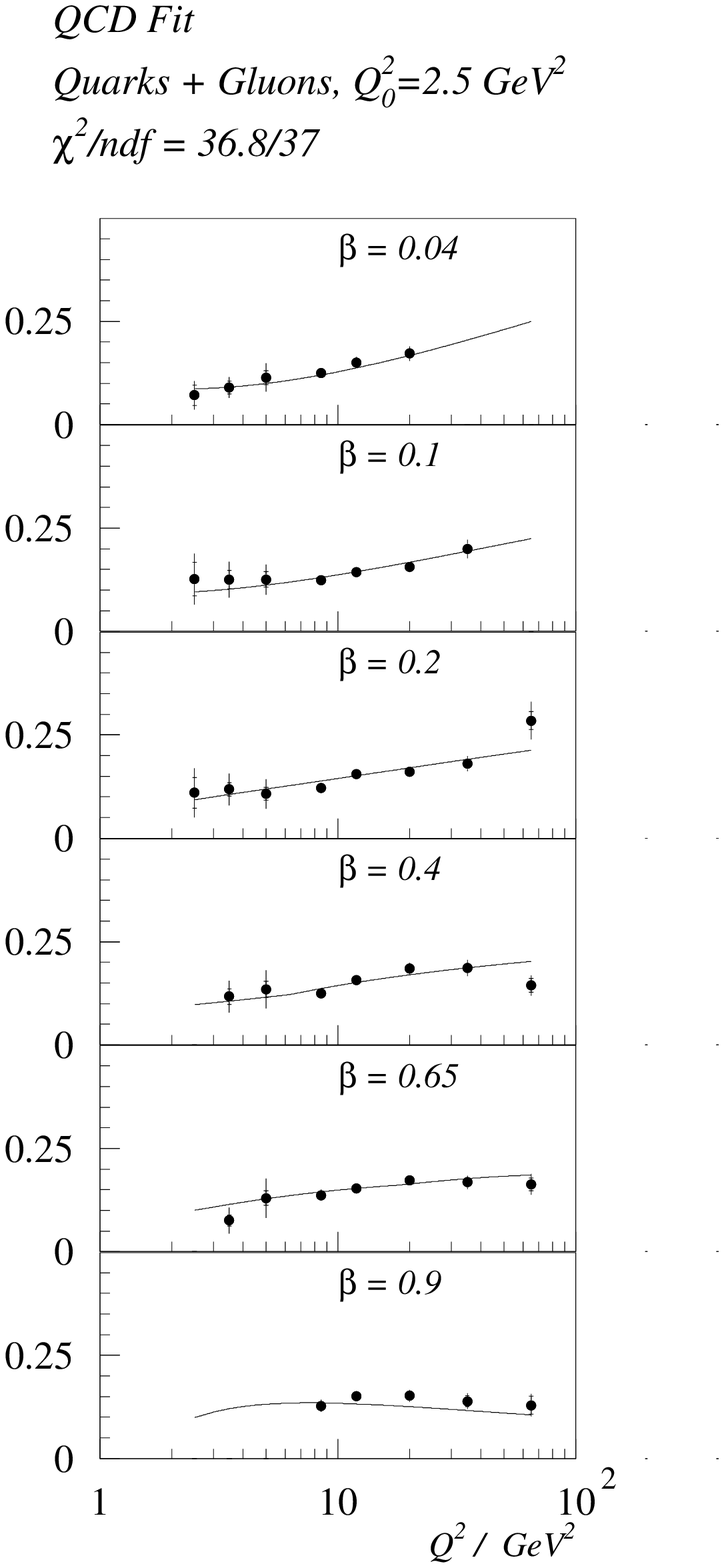,height=9cm,width=6cm,%
bbllx=0pt,bblly=90pt,bburx=290pt,bbury=685pt,clip=}
\vspace{-0.5cm}
\caption\protect{The $Q^2$ dependence of $\tilde{F}_2^D \prop F_2^{\pom}$ 
in bins of $\beta$. The lines are the result of a QCD fit with quarks and
gluons present at the starting scale of $Q^2_0=2.5 \gev2$.}
\label{fig:f2pom}
\end{figure}

The $Q^2$ dependence in bins of $\beta$ of $\tilde{F}_2^D=\int
F_2^{D(3)} dx_{\pom}$, where the integration is over $3\!\cdot\!10^{-4}
\!<\!x_{\pom}\!<\! 5\!\cdot\!10^{-2}$, as determined by the H1 
experiment is presented in figure~\ref{fig:f2pom}.
\begin{figure}[hbt]
\center
\epsfig{figure=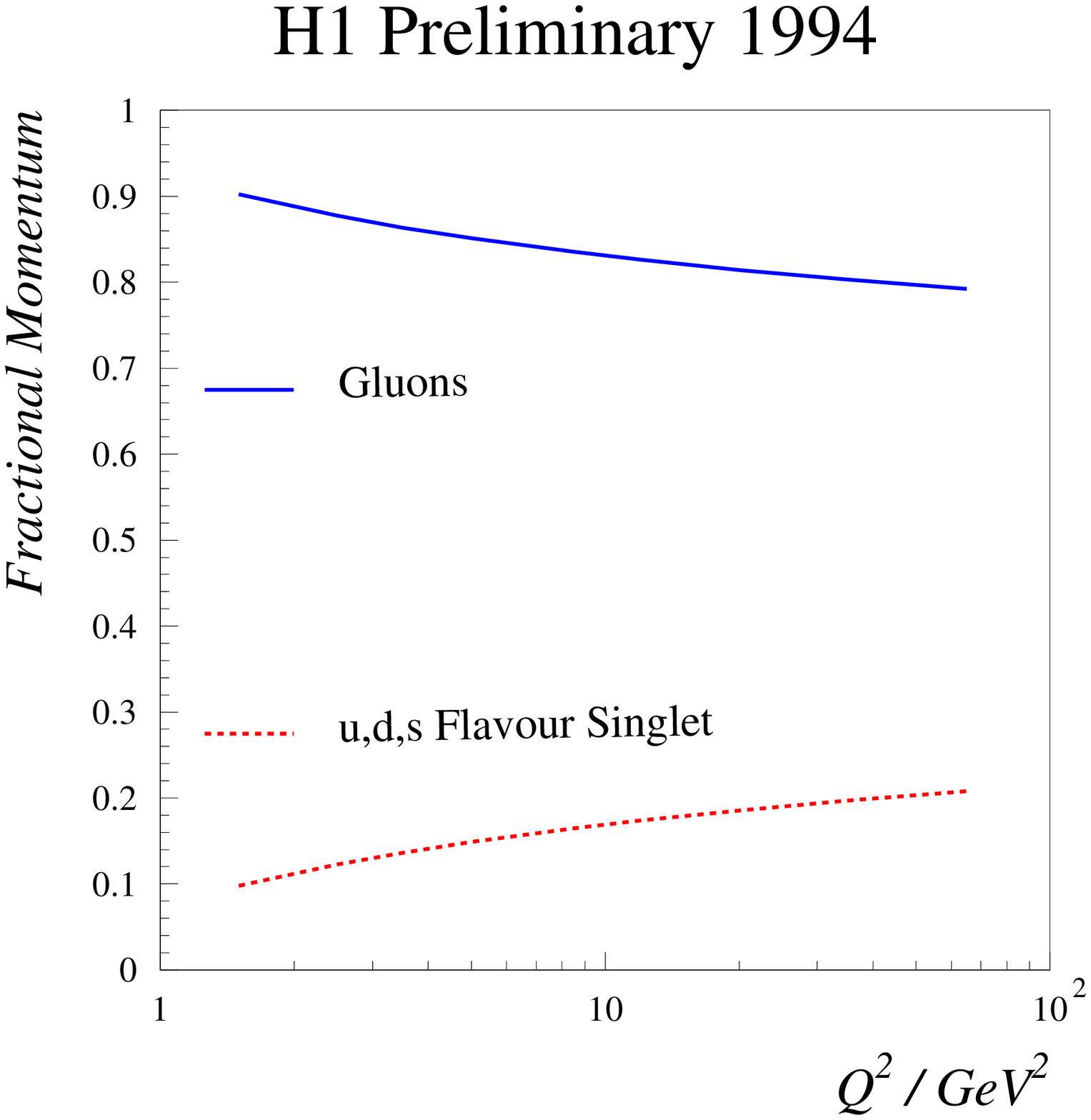,width=5cm}
\vspace{-0.5cm}
\caption\protect{The fraction of the $\pom$ momentum carried by gluons 
and light quarks as a function of $Q^2$ as determined from a QCD fit.
}
\label{fig:sumrule_lrg}
\end{figure}
While the $\beta$ dependence is very weak, the striking observation is
that at large $\beta$ there is almost no $Q^2$ dependence. A slight
rise with $Q^2$ is evident at low $\beta$. The structure function
$\tilde{F}_2^D$ is directly proportional to $F_2^{\pom}$. If Regge
factorization holds, one would expect the $Q^2$ dependence of
$F_2^{\pom}$ to be governed by the DGLAP evolution equations. In this
case the lack of $Q^2$ dependence at large $\beta$ can only be
explained by a large gluon contribution at large $\beta$ which would
compensate the expected depletion of fast quarks due to
radiation. Indeed the initial conditions needed to fulfill the NLO QCD
evolution of $F_2^{\pom}$, as obtained from a fit performed by H1,
favor a gluon distribution strongly peaked at fraction of $\pom$
momentum close to unity and a flat quark distribution.  As seen in
figure~\ref{fig:sumrule_lrg} more than $80\%$ of the Pomeron momentum
is carried by (hard) gluons, at least in the $Q^2$ range available in
this study.

\subsection{DIS vector meson production}

Another diffractive reaction which was investigated at HERA is the
exclusive vector meson (VM) production in DIS, $ep \rightarrow
epV$. The interest in measuring this process was generated by the
calculations of VM production in two-gluon exchange
models.\cite{DL_rho,ryskin_jpsi,Brod94} Two-gluon exchange is a
prototype for modeling the Pomeron exchange.\cite{Low,Nussinov} It
turns out that for a longitudinally polarized photon perturbative QCD
calculations can be applied strictly and the applicability of the QCD
factorization theorem implies that the scattering amplitude is
proportional to the gluon distribution in the
proton.\cite{Brod94,koepf} The contribution of the longitudinal
polarization is expected to dominate at large $Q^2$, independently of
other model assumptions. Thus in all the models, at small $x$
\be
\frac{d \sigma}{dt} \propto \frac{\alpha_s^2(Q^2)}{Q^6} |xg(x,Q^2)|^2 \, ,
\label{eq:sigma_vm}
\ee
however the nature of the exchanged gluons in the VM production
mechanism is subject to heated debates (for a review
see~\cite{hera_ws96_vm}).

Should the QCD expectations be fulfilled in the data, hard diffractive
VM production may turn out to be the way to measure the gluon
distribution in the proton, to study the approach to the unitarity
limit and even to investigate the properties of the VM wave functions.
\begin{figure}[hbt]
\center
\vspace{-0.3cm}
\epsfig{figure=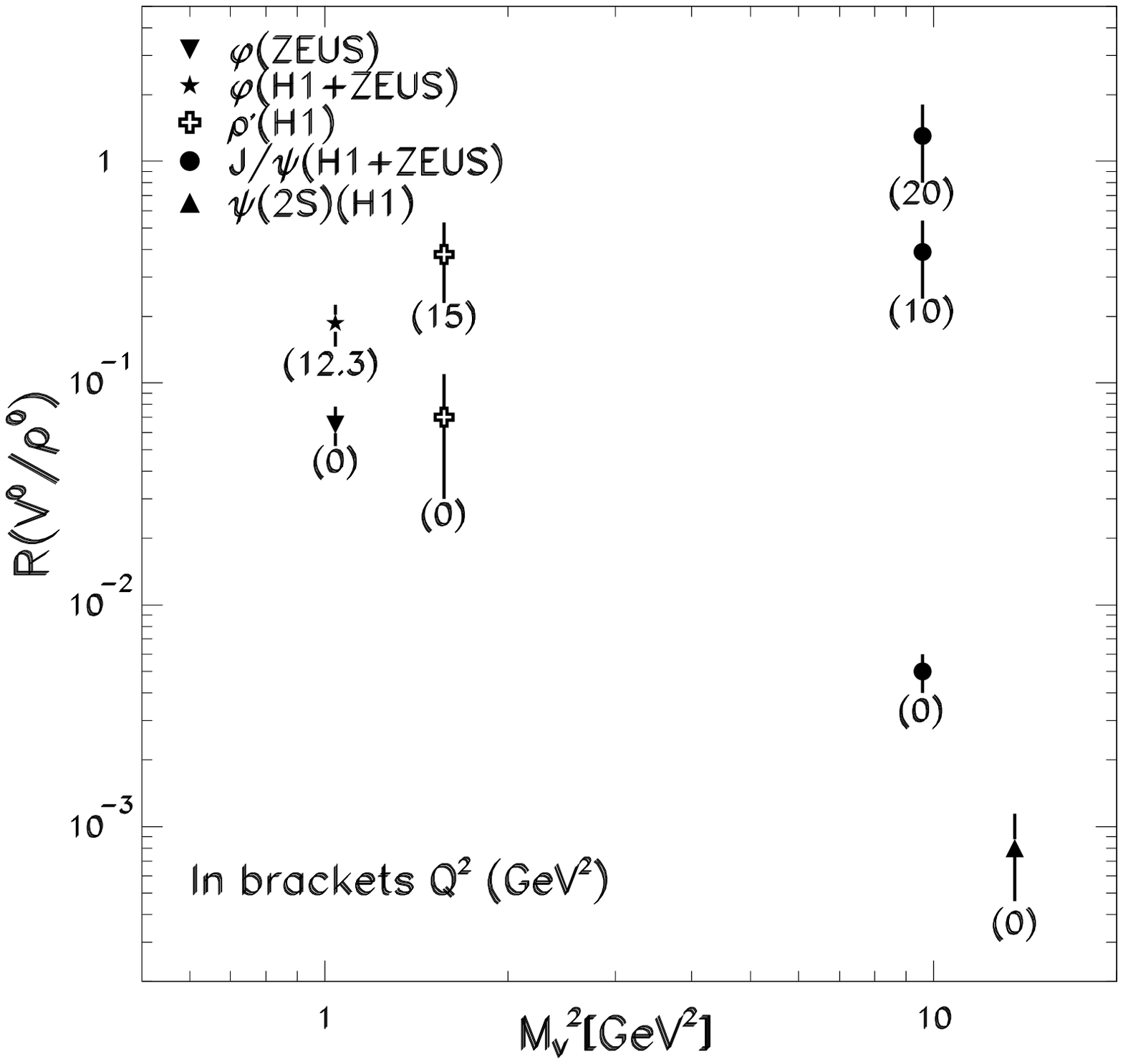,width=6.5cm}
\vspace{-0.3cm}
\caption\protect{Production rate of  
vector mesons relative to $\rho^0$ as a function of their mass square
$M_V^2$ for different values of $Q^2$. }
\label{fig:su4}
\vspace{-0.5cm}
\end{figure}
There are indications that the perturbative production mechanism
is present. The measurements of the $Q^{-2a}$ dependence of the
$\rho^0$ and $J/\psi$ production cross sections favor a value of $a
\simeq 2$ with an error typically of $20
\%$.\cite{h1_dis_rhopsi,zeus_dis_vm} In photoproduction the SU(4) flavor
symmetry is badly broken. In QCD one expects the SU(4) symmetry to be
restored at large $Q^2$. This seems to be borne out by the data as can
be seen in figure~\ref{fig:su4} where the measured rates of various VM
production relative to the $\rho^0$ are shown at $Q^2=0 \gev2$ and for
higher $Q^2$.\cite{h1_dis_rhopsi,zeus_dis_vm,h1_dis_vm,ZEUS_phi} The
most spectacular is the increase of the $J/\psi$ to $\rho^{0}$
production ratio which approaches unity at $Q^2=20 \gev2$.

\begin{figure}[hbt]
\center
\epsfig{figure=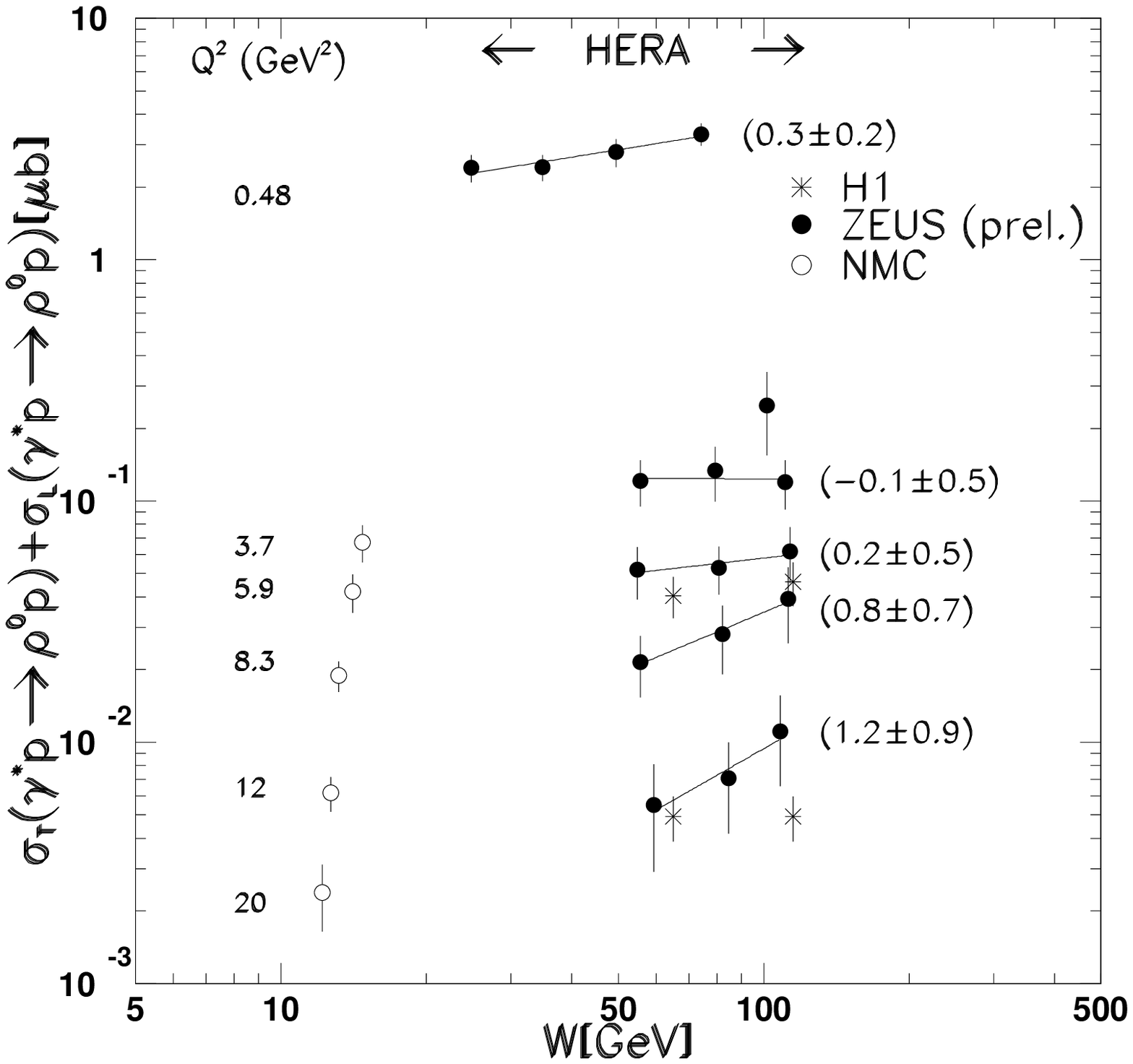,width=6.5cm}
\vspace{-0.3cm}
\caption\protect{The $\rho^0$ cross section as 
a function of $W$ for different $\gamma^*$ virtualities $Q^2$. The
numbers in parenthesis are the values of the power $a$ of a fit $W^a$
to the ZEUS data.}
\label{fig:xsec_vm}
\end{figure}

Should the gluons mediating the VM production be perturbative in
nature, one expects the cross section to increase with $W$ as the
square of the gluon distribution, that is, much faster than for the
soft Pomeron. At high $Q^2$ the HERA measurements are not precise
enough to determine the $W$ dependence consistently within one
experiment. Relative to the NMC measurements~\cite{nmc_vm} the rise of
the cross section is much stronger than expected for non-perturbative
production mechanisms. Also the $W$ dependence of the $J/\psi$
photoproduction cross section is stronger than for the soft
Pomeron~\cite{h1_ph_jpsi,zeus_ph_jpsi} and in accord with the
perturbative approach.\cite{ryskin_jpsi,durham_jpsi} A compilation of
the existing measurements of $\rho^0$ is presented in
figure~\ref{fig:xsec_vm}.

\begin{figure}[htb]
\center
\epsfig{figure=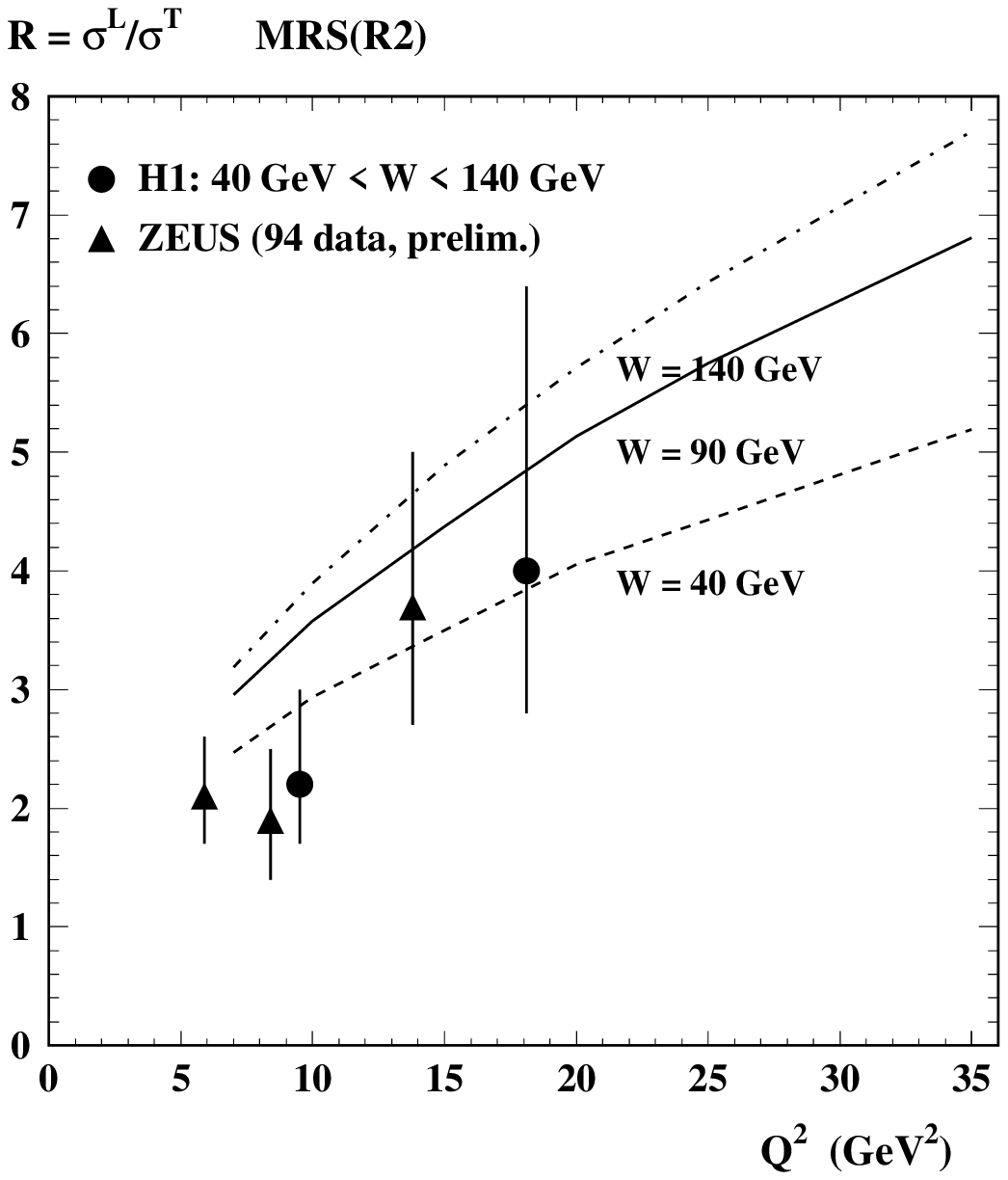,width=6cm}
\vspace{-0.5cm}
\caption\protect{The ratio $R$ of longitudinal to transverse $\rho^0$ 
polarization as a function of $Q^2$ compared to perturbative model
calculations.\cite{durham_rho}}
\label{fig:R_rho}
\end{figure}

\begin{figure}[htb]
\center
\epsfig{figure=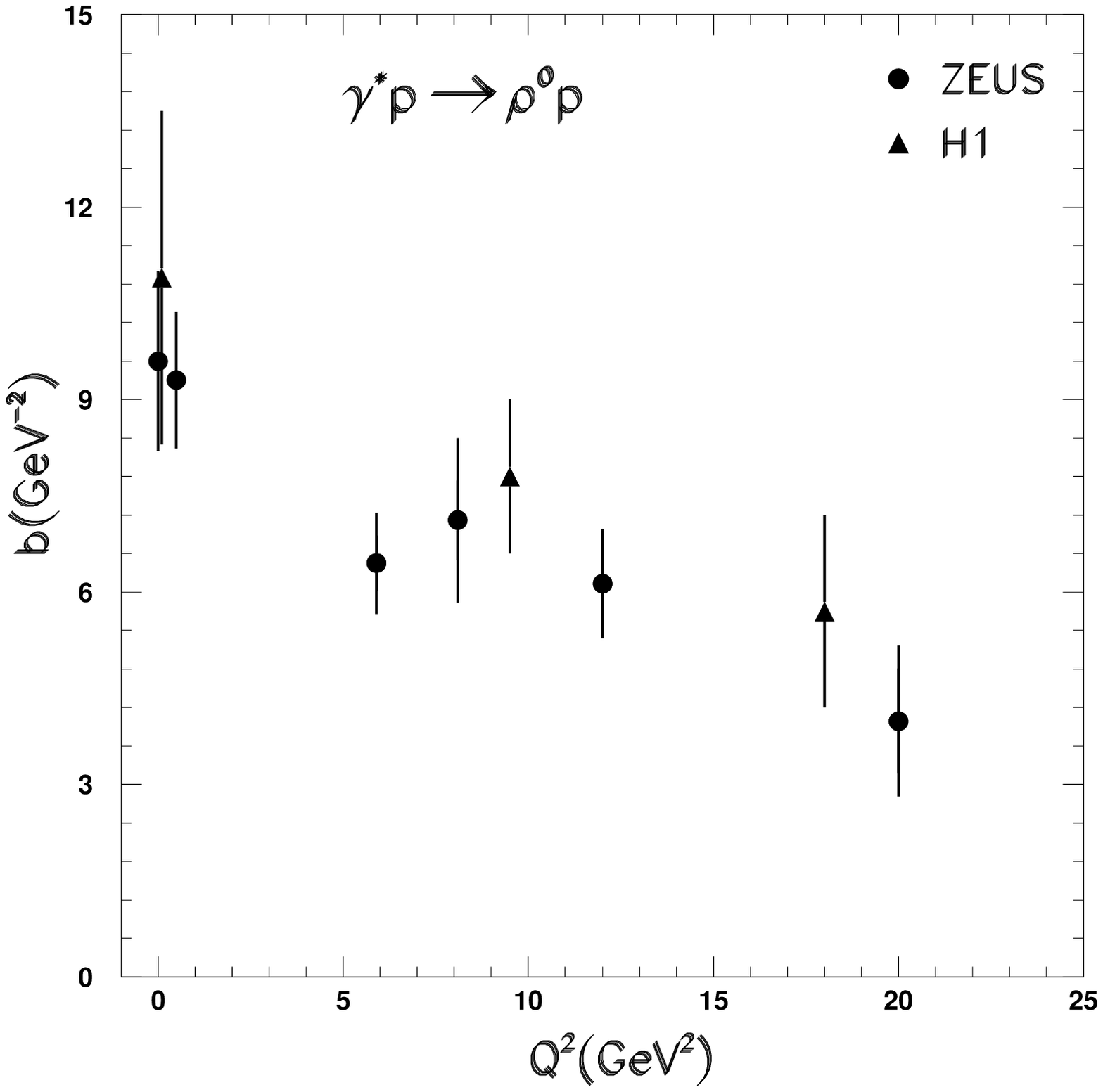,width=6.5cm}
\vspace{-0.3cm}
\caption\protect{The slope $b$ of the exponential $t$ dependence 
$\exp(-b|t|)$ for $\rho^0$ production as a function of the photon
virtuality $Q^2$.}
\label{fig:bslope_q2}
\end{figure}

\begin{figure}[htb]
\center
\epsfig{figure=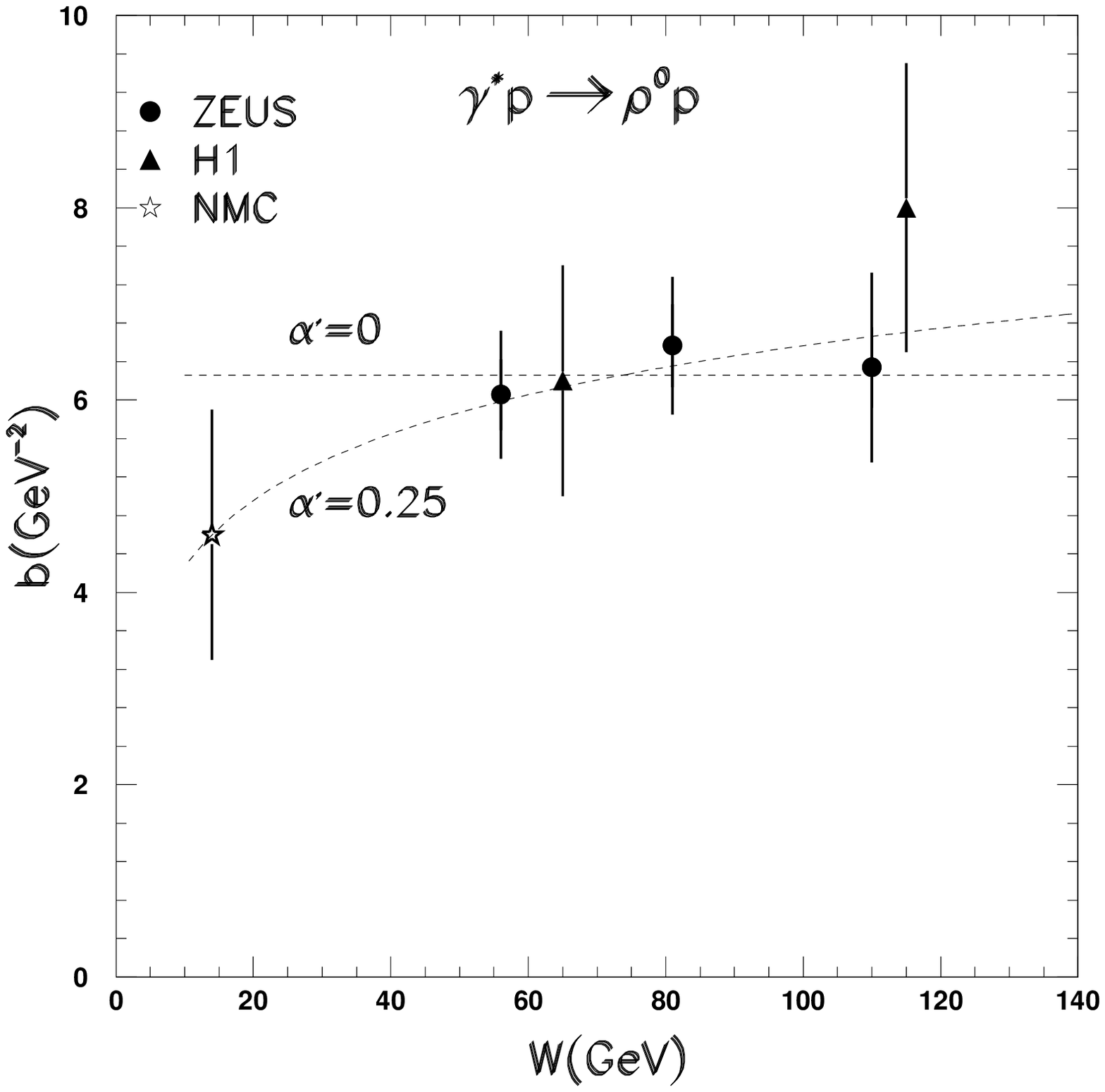,width=6.5cm}
\vspace{-0.3cm}
\caption\protect{The slope $b$ of the exponential $t$ dependence 
$\exp(-b|t|)$ for $\rho^0$ production at $Q^2 \simeq 10 \gev2$ as a
function of $W$.}
\label{fig:bslope_w}
\end{figure}

The contribution of the longitudinal photon relative to the transverse
photon, $R$, can be determined from the polarization of the vector meson
and assuming s-channel helicity conservation. As expected $R$ is
increasing with $Q^2$, as shown in figure~\ref{fig:R_rho}, and this
increase is well reproduced by perturbative calculations which assume
hadron-parton duality.\cite{durham_rho}
The decrease of the size of the interacting photon is also observed in
the behavior of the slope $b$ of the $t$ distribution. This can be
seen in figure~\ref{fig:bslope_q2} where $b$ measured for exclusive
$\rho^0$ production is plotted as a function of $Q^2$. The $b$
dependence as a function of $W$ in DIS is compatible with $\alpha' =
0.25$~GeV$^{-2}$ as in the case of reactions mediated by the soft
Pomeron (see figure~\ref{fig:bslope_w}).

\subsection{Conclusions}

The intercept of the Pomeron as measured in DIS seems higher than the
one of the Pomeron mediating soft interactions. At this point it is
not clear whether we are probing the same object and it might be wiser
to use the notion of an effective Pomeron. Assuming the validity of
the DGLAP evolution for inclusive diffractive scattering, it is found
that the effective Pomeron consists mainly of hard gluons. There are
some indications that it develops differently in exclusive processes
when $M_X^2 \ll Q^2$. We may be witnessing the fact that the Pomeron
is not a universal object. However, before this conclusion can be
reached on solid grounds more precise measurements are needed.

\section{The photon structure function}

The photon is probably the most interesting particle to study. It is
one of the gauge particles of the Standard Model and as such has no
intrinsic structure. However it acquires a structure in its
interactions with matter and in that sense it is a prototype for
studying the formation of a partonic object. 

The structure functions of the photon are defined in DIS $e \gamma
\rightarrow e X$ through the interaction cross section,
\be
\frac{d^2\sigma}{dxdQ^2}=\frac{2\pi\alpha^2}{xQ^4}\left[ \left( 1+(1-y)^2
\right) F_2^{\gamma} - y^2F_L^{\gamma} \right] \, .
\label{eq:xsec_dis_ph}
\ee
In the quark-parton model the contribution to this cross section comes
from the so called box diagram $\gamma^* \gamma \rightarrow q
\bar{q}$. Should this be the only contribution, the structure function of 
the photon could be fully calculated in QCD. In reality the real
photon can fluctuate into partonic states and in particular into a
vector meson state. Thus the overall contribution to the cross section
is more complicated. This has also implications for the evolution
equation of the photon structure function. In particular the box
contribution mentioned above introduces an extra splitting function
into the evolution equation, that of a photon splitting into a $q
\bar{q}$. The latter is responsible for the fact that 
$F_2^{\gamma}(x,Q^2)$ is large at large $x$ and increases with $Q^2$
at any value of $x$. 

Experimentally $F_2^{\gamma}$ is measured in $e^+e^-$
interactions, by requiring that one of the electrons scatters under
small angles and remains undetected (the source of the target photon)
while the second one scatters under a large angle, providing the
probing virtual photon. New data have been
presented~\cite{OPAL,DELPHI} on the measurements of the $\gamma^*
\gamma$ interactions from the LEP experiments. The OPAL 
experiment~\cite{OPAL} has presented results on $F_2^{\gamma}$ with
improved resolution, thanks to the use of their forward detectors.
\begin{figure}[htb]
\center
\epsfig{figure=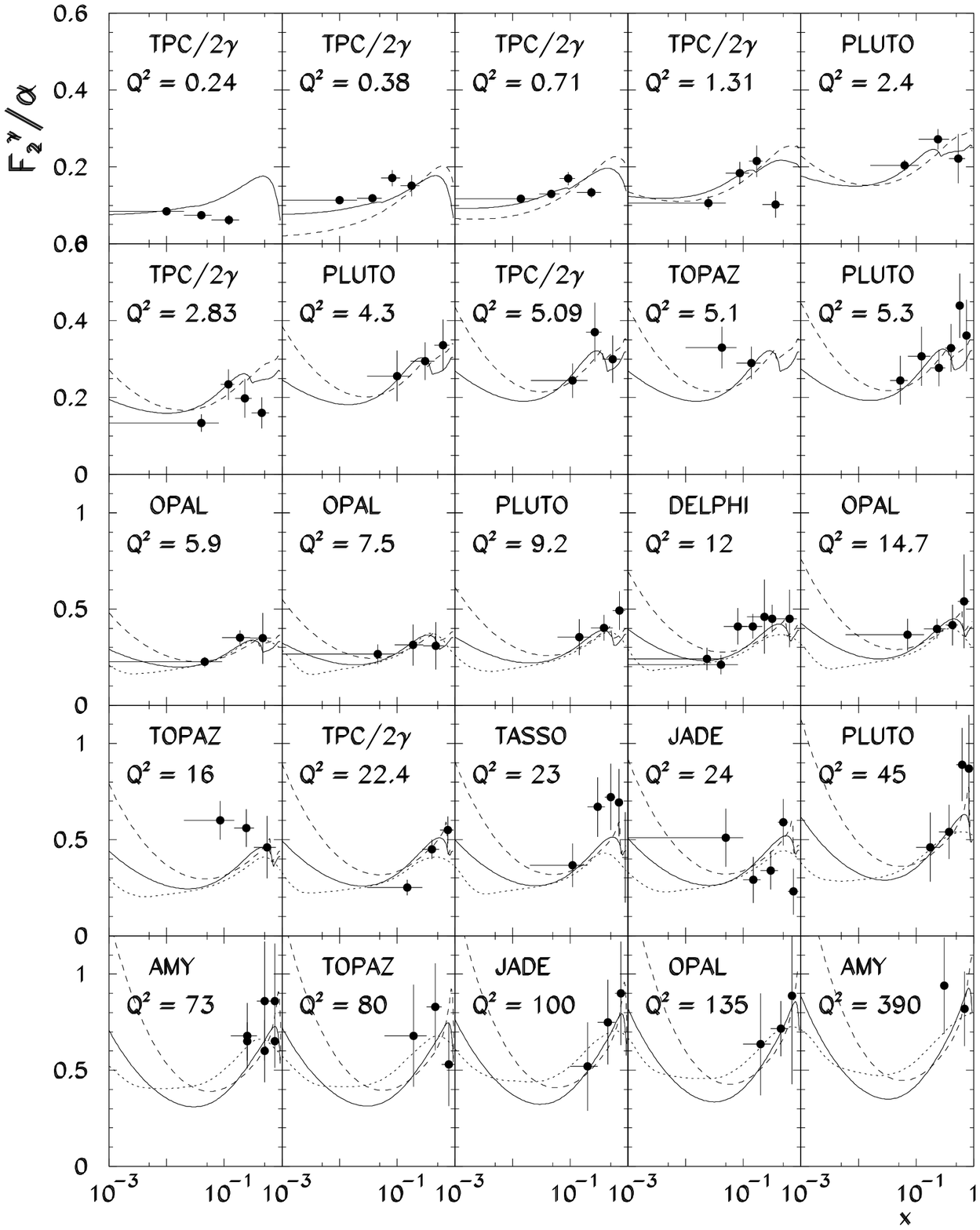,width=8cm}
\caption\protect{The photon structure function $F_2^{\gamma}/\alpha$ 
as a function of $x$ in bins of $Q^2$ compared to the LO
GRV~\cite{grv_f2ph} (dashed line), GS~\cite{gs_f2ph} (dotted line) and
SaS~\cite{sas_f2ph} (full line) parameterizations of parton
distributions in the photon.}
\label{fig:f2g}
\end{figure}

The $x$ dependence of $F_2^{\gamma}$ in bins of $Q^2$ is shown in
figure~\ref{fig:f2g}.  For the sake of comparison with the proton
structure function the scale of $x$ was chosen logarithmic. The
measurements are compared to three different sets of parton
distributions in the photon.\cite{grv_f2ph,gs_f2ph,sas_f2ph} At higher
$x$, where data are available, all the parameterizations describe
$F_2^{\gamma}$ well. The largest differences are seen at low $x$. This
is not unlike the parameterizations of the proton structure function
which were available before the low $x$ measurements at HERA. In the
case of the photon, the situation is thought to be worse since no
momentum sum rule is used to constrain the gluon distribution. However
recently such a sum rule has been derived.\cite{gamma_sumrule}

\begin{figure}[htb]
\center
\epsfig{figure=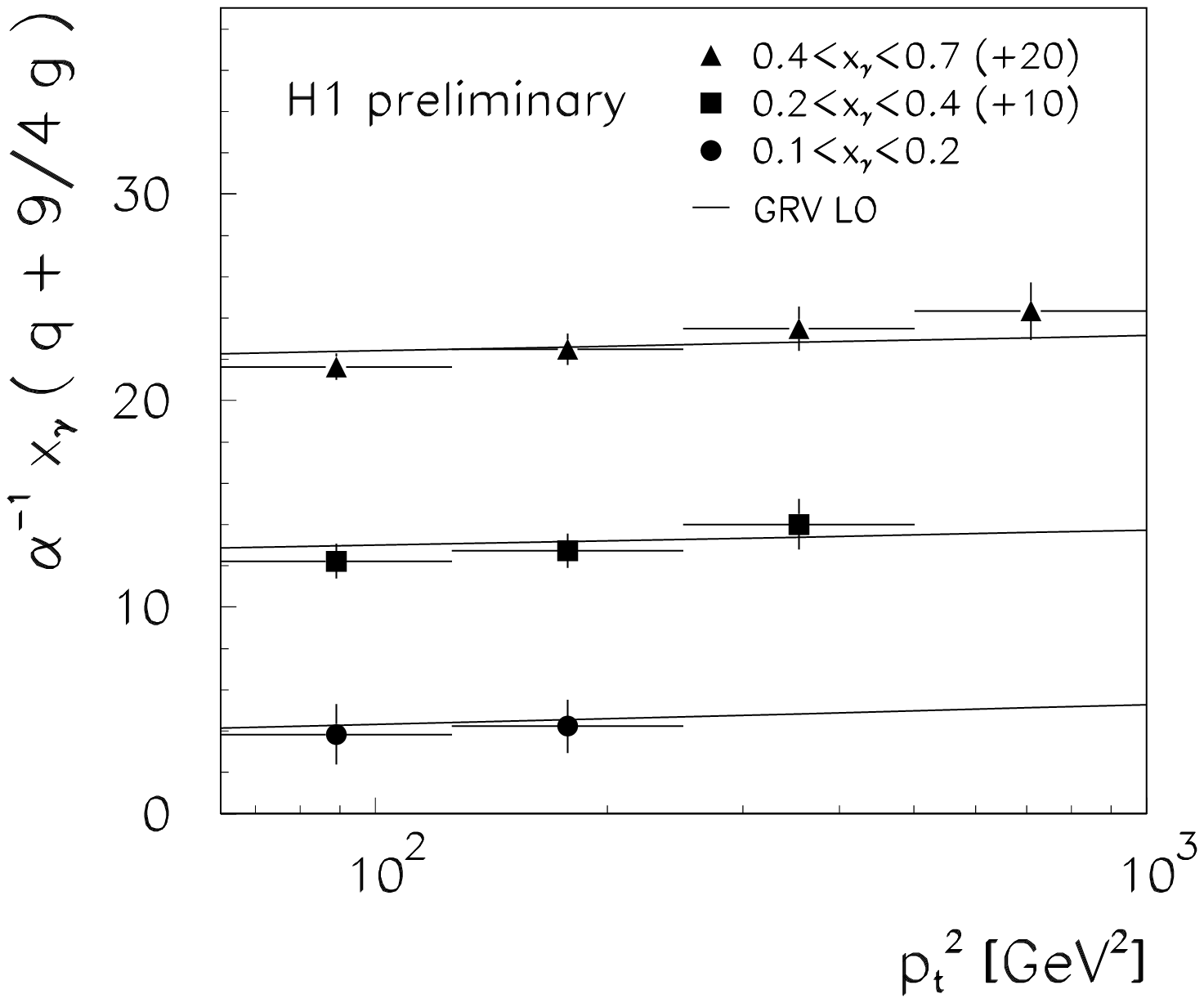,width=7cm}
\vspace{-0.5cm}
\caption\protect{Effective structure function of the photon as measured 
in photoproduction of two jets with high transverse momentum $p_t$.}
\label{fig:h1_photon}
\end{figure}

The uncertainty in the gluon distribution can be partly minimized by
the hard photoproduction data from the HERA
experiments~\cite{h1_f2gamma} $\gamma p
\rightarrow jet_1+jet_2+X$. An example of the potential of HERA is
shown in figure~\ref{fig:h1_photon}, where an effective structure
function of the photon is plotted as a function of the transverse
momentum square of the jets, for various ranges of $x$ of the photon.

\section{Summary}

The $F_2$ structure function of the proton has been measured with high
precision in the range $10^{-6}\!<\!x\!<\!1$ and $0.1\!<\!Q^2\!<\!5000
\gev2$. The DGLAP evolution equation in NLO describes the data very
consistently and down to $Q^2 \simeq 1 \gev2$. However the
contribution of non-conventional QCD dynamics cannot be ruled out
without further studies.

Hard diffractive scattering, first sign of coherent phenomena in
perturbative QCD, allows to probe directly the nature of strong
interactions. It can further help to constrain the gluon distribution
in the proton.

With the LEP and HERA experiments measuring the partonic structure of
the photon we are slowly approaching the low $x$ regime in photon
induced reactions.

Given the large theoretical support and interest in the low $x$
phenomena, we should soon be able to achieve a good understanding of
QCD dynamics of high parton densities. One of the outcomes will be
reliable parton distributions for the next generation colliders.

\section*{Acknowledgments}

It is a pleasure for me to thank all my colleagues from the different
experiments for useful discussions. Very special thanks go to
E. Gurvich, A. Levy and E. Rondio for preparing some of the
compilation figures.  I am grateful for the help of R. Klanner,
A. D. Martin and R. G. Roberts in preparing this talk. I would like to
acknowledge many enlightening discussions with J. Bartels,
W. Buchm\"uller and L. Frankfurt.  Last but not least I would like to
thank the organizers and colleagues from Warsaw University (my Alma
Mater) for their warm hospitality.

This work was partially supported by the German Israeli Foundation,
the Minerva Foundation and DESY.

\section*{References}
%\bibliography{proc_b}

\end{document}